\documentclass{aastex61}

\usepackage{amssymb,amsmath,amsfonts,amsbsy,graphicx,bm}
\usepackage{ulem}

\newcommand{\calC}{{\cal C}}
\newcommand{\calD}{{\cal D}}
\newcommand{\calO}{{\cal O}}
\newcommand{\calR}{{\cal R}}

\begin{document}

\title{Analytic integral solutions for induced gravitational waves}

\author{Jinn-Ouk Gong}

\affiliation{Korea Astronomy and Space Science Institute, Daejeon 34055, Korea}
\affiliation{Department of Science Education, Ewha Womans University, Seoul 03760, Korea}
\affiliation{Asia Pacific Center for Theoretical Physics, Pohang 37673, Korea}

\begin{abstract}

We present analytic integral solutions for the second-order induced gravitational waves (GWs). After presenting all the possible second-order source terms, we calculate explicitly the solutions for the GWs induced by the linear scalar and tensor perturbations during matter- and radiation-dominated epochs.

\end{abstract}

\keywords{cosmology: theory --- large-scale structure of universe --- gravitational waves}

\newpage

\section{Introduction}
\label{sec:intro}


A series of detection of the gravitational waves (GWs) by the LIGO and Virgo collaborations~\citep{2016PhRvL.116f1102A, 2016PhRvL.116x1103A, 2017PhRvL.118v1101A, 2017PhRvL.119n1101A, 2017PhRvL.119p1101A, 2017ApJ...851L..35A} has opened the era of multi-messenger astronomy led by GWs. The observed GW signals are originated from merging black holes and/or neutron stars, but there should be other events energetic enough to generate observable GWs. Such cosmological origins include cosmic strings~\citep{1985PhRvD..31.3052V,1986NuPhB.277..605B}, phase transition~\citep{1984PhRvD..30..272W,1986MNRAS.218..629H} and preheating~\citep{1997PhRvD..56..653K}. But the non-linear nature of gravity tells us that there is a persistent source of GWs -- they can be induced by other cosmological perturbations at non-linear order~\citep{2004PhRvD..69f3002M,2007PhRvD..75l3518A,2007PhRvD..76h4019B}: while the scalar-, vector- and tensor-type perturbations are decoupled at linear order, they couple to each other and thus can be generated at non-linear order~\citep{2004PhRvD..69j4011N,2007PhRvD..76j3527H}. Especially, the second-order tensor perturbations, or GWs, induced by linear scalar perturbations\footnote{
It should be noted that large scalar perturbations may well be induced by large tensor perturbations on small scales, leading to copious production of primordial black holes~\citep{2015PhRvD..92l1304N,2016PhRvD..94d3507N}. Such induced scalar perturbations subsequently can further induce tensor perturbations. This is an interesting possibility, but is beyond the scope of the present work.
} may well be sizable if on small scales the primordial curvature perturbation is enhanced during inflation~\citep{2012JCAP...09..017A} or the density perturbation grows during an early matter-dominated (MD) epoch~\citep{2009PhRvD..79h3511A,2010JCAP...04..021J,2013JCAP...05..033A}.


In most of the literature, however, the study of second-order induced GWs has been focused on MD in one particular gauge choice for the scalar perturbations. The reason could be twofold. First, the (geometric) scalar perturbation remains constant even on sub-horizon scales during MD, but it decays quickly if it enters the horizon during radiation-dominated (RD) epoch [see \eqref{eq:scalarRD-ZSG} and \eqref{eq:scalarRD-com}]. Thus, even if the primordial curvature perturbation is enhanced on small scales, one would naively expect such an enhancement to disappear during an RD epoch so that there should be no sizable induced GWs from scalar perturbations [see, however,~\cite{2019JCAP...10..071I,2019PhRvD.100d3532I}]. Furthermore, there is an upper bound on the contribution of the primordial GWs from the observations on the cosmic microwave background (CMB): in terms of the so-called tensor-to-scalar ratio, $r < 0.07$ at the pivot scale $k=0.05$ Mpc$^{-1}$~\citep{2016PhRvL.116c1302B}. This means on the CMB scales the amplitude of the primordial GWs should be about 1/10 or even smaller than that of the primordial curvature perturbation. Thus, naturally scalar perturbations should be the most dominant source for the induced GWs compared to the other types of cosmological perturbations. For these reasons, it is very sensible to consider the induced GWs by scalar perturbations during MD.


Nevertheless, this does not mean at all that we have a complete and satisfactory understanding of the second-order induced GWs. First, there is no a priori reason why scalar perturbations should be considered only in a particular gauge condition. Moreover, as the linear scalar perturbations depend on the choice of gauge, the second-order GWs induced by their quadratic combinations should be also dependent on the gauge conditions. This is obvious and indeed was noticed early~\citep{2009PhRvD..80l3526A} but was explicitly shown only recently in~\cite{2017ApJ...842...46H}. Second, while scalar perturbations are likely to be the most important source for the induced GWs, other types of perturbations need not be neglected from the beginning. Especially, the linear tensor perturbations should be persistent. Furthermore, it is possible that the contributions of the tensor perturbations can be enhanced~\citep{2006JCAP...02..004M}, thus, the induced second-order GWs from them may well be significant accordingly\footnote{
More exactly, if the tensor-induced GWs are to be dominant, the scalar perturbations should not increase. Otherwise, the scalar-induced GWs are very likely to be more prominent than the tensor-induced ones. This condition allows certain models viable [e.g.~\cite{2009JCAP...01..009J, 2010PhRvD..82b3509J}], but generally excludes models that include a break during inflation [e.g.~\cite{2019JCAP...06..049P}].
}. This indeed happens in certain concrete models beyond the standard slow-roll inflation~\citep{2007PhRvL..98w1302B,2010PhRvL.105w1302K,2014JCAP...07..022G, 2015NuPhB.900..517C,2018JCAP...12..024M}. Third, as the universe has evolved through both RD and MD, for a complete description of the induced GWs we need a proper understanding of RD as it is occurring. This was recognized early as well~\citep{2007PhRvD..75l3518A,2010PhRvD..81b3527A}, but analytic approach has been taken only recently~\citep{2018JCAP...09..012E,2018PhRvD..97l3532K}.

In this article, we provide analytic integral solutions for the second-order induced GWs from both linear scalar and tensor perturbations. We also present the full second-order source terms with all three types of cosmological perturbations, so it should be straightforward to calculate the solutions from the sources with vector perturbations. This article is outlined as follows. In Section~\ref{sec:equation}, we provide the full traceless evolution equation for the spatial metric tensor, including all the explicit second-order source terms. In Section~\ref{sec:linearsol}, we solve the equations of motion for the linear cosmological perturbations that will be used in Section~\ref{sec:inducedGWsols} to compute the analytic integral solutions for the second-order induced GWs. We briefly summarize our results in Section~\ref{sec:conc}. Some technical details are relegated to the appendix sections.

\section{Second-order equation}
\label{sec:equation}

Our metric convention of a flat Friedmann universe including cosmological perturbations is
\begin{equation}
\label{eq:metric}
ds^2 
= 
-a^2 (1+2\alpha) d\eta^2 - 2a^2 B_i d\eta dx^i
+ a^2 \Big[ (1+2\varphi) \delta_{ij} + 2\gamma_{,ij} + 2C_{(i,j)} + 2h_{ij} \Big]
dx^i dx^j
\, ,
\end{equation}
where $d\eta = dt/a$ is the conformal time and $a(\eta)$ is the scale factor. The indices of the perturbation variables are raised and lowered by $\delta_{ij}$. Further, the shear $\chi_i$ is written as
\begin{equation}
\chi_i 
\equiv
a \Big( B_i + a\dot\gamma_{,i} + a\dot{C}_i \Big)
\, .
\end{equation}

The non-linear equation necessary for the induced GWs can be obtained from the traceless evolution equation for the spatial metric. Up to second order, the full equation is given by \eqref{eq:traceless-eq} in Appendix~\ref{app:traceless-eq}. Further, writing $B_i = \beta_{,i} + B^{(v)}_i$ with $B^{(v)i}{}_{,i} = 0$, the shear $\chi_i$, anisotropic stress $\Pi_{ij}$ and peculiar velocity $v_i$ of the perfect fluid, whose energy density and pressure are written as $\rho$ and $p$ respectively, can be decomposed in terms of the scalar gradient, transverse vector and transverse and traceless tensor components as
\begin{align}
\chi_i 
&
=
a \big( \beta + a\dot\gamma \big)_{,i}
+ a \Big( B^{(v)}_i + a\dot{C}_i \Big)
\equiv
\chi_{,i} + \chi_i^{(v)}
\, ,
\\
\Pi_{ij} 
& 
= 
\frac{1}{a^2} \bigg( \Pi_{,ij} - \frac{\delta_{ij}}{3}\Delta\Pi \bigg)
+ \frac{1}{a}\Pi^{(v)}_{(i,j)} + \Pi^{(t)}_{ij} + \frac{\delta_{ij}}{3}\Pi^k{}_k
\, ,
\\
v_i
& 
=
-v_{,i} + v^{(v)}_i
\, ,
\end{align}
where obviously the superscripts $(v)$ and $(t)$ denote respectively the transverse vector and transverse and traceless tensor. Note that the last term of $\Pi_{ij}$ is added because $\delta^{ij}\Pi_{ij} \equiv \Pi^i{}_i \neq 0$ at non-linear order but is given by
\begin{equation}
\Pi^i{}_i
=
2h^{ij}\Pi_{ij} + v^iv^j\Pi_{ij} + \cdots
\end{equation}

With these decompositions, the second-order traceless evolution equation \eqref{eq:traceless-eq} can be written as
\begin{equation}
\label{eq:traceless-eq2}
\ddot{h}_{ij} + 3H\dot{h}_{ij} - \frac{\Delta}{a^2}h_{ij} - 8\pi G\Pi^{(t)}_{ij}
+ \frac{1}{a^2} \bigg( \partial_i\partial_j - \frac{\delta_{ij}}{3}\Delta \bigg) 
\bigg[ \frac{1}{a}\frac{d}{dt} \big( a\chi \big) - \alpha - \varphi - 8\pi G\Pi \bigg]
+ \frac{1}{a} \bigg[ \frac{1}{a^2} \frac{d}{dt} \Big( a\chi^{(v)}_{(i,j)} \Big) 
- 8\pi G \Pi^{(v)}_{(i,j)} \bigg]
=
s_{ij} 
\, ,
\end{equation}
where $s_{ij}$ denotes all second-order terms, which serve as the second-order ``source''. As there are scalar, vector and tensor perturbations, at second order 6 combinations are possible:

1. Scalar-scalar source $s_{ij}^{(ss)}$ denotes the collection of the products of two scalar perturbations:
\begin{align}
s_{ij}^{(ss)}
=
&
\frac{1}{a^3} \frac{d}{dt} \Big[
a \Big( 2\varphi\chi_{,ij} + \varphi_{,i}\chi_{,j} + \varphi_{,j}\chi_{,i} \Big)
\Big]
+ \frac{1}{a^2} \Big(
\kappa\chi_{,ij} - 4\varphi\varphi_{,ij}  -   3\varphi_{,i}\varphi_{,j}
\Big)
+ \frac{1}{a^4}\chi^{,k}{}_{,i}\chi_{,jk}
\nonumber\\
& 
+ \frac{1}{a^2}  \Big[ 2\alpha\dot\chi_{,ij} - H\alpha\chi_{,ij}  + \dot\alpha\chi_{,ij} 
- 2(\alpha+\varphi)\alpha_{,ij} - \alpha_{,i}\alpha_{,j} - 2\varphi_{(,i}\alpha_{,j)}
\Big]
+ 8\pi G (\rho+p) v_{,i}v_{,j} - 16\pi G \varphi  \frac{\Pi_{,ij}}{a^2} 
\nonumber\\
& - \frac{\delta_{ij}}{3} \bigg\{ 
\frac{1}{a^3} \frac{d}{dt} \Big[ 
a \Big( 2\varphi\Delta\chi + 2\varphi_{,k}\chi^{,k} \Big)
\Big]
+ \frac{1}{a^2}  \Big(
\kappa\Delta\chi - 4\varphi\Delta\varphi   - 3\varphi^{,k}\varphi_{,k}
\Big)
+ \frac{1}{a^4} \chi^{,kl}\chi_{,kl}
\nonumber\\
& \qquad 
+ \frac{1}{a^2}  \Big[ 
2\alpha\Delta\dot\chi - H\alpha\Delta\chi + \dot\alpha\Delta\chi 
- 2(\alpha+\varphi)\Delta\alpha - \alpha^{,k}\alpha_{,k} - 2\alpha^{,k}\varphi_{,k}
\Big]
+ 8 \pi G (\rho+p) v^{,k}v_{,k} - 16\pi G\varphi\frac{\Delta}{a^2}\Pi \bigg\}
\, .
\end{align}

2. Scalar-tensor source $s_{ij}^{(st)}$ denotes the collection of the products of one of the scalar perturbations and tensor perturbations:
\begin{align}
s_{ij}^{(st)} =
& \frac{d}{dt} \bigg[ 
 \dot{h}_{ij}   \alpha
+ 2 \bigg( \varphi\dot{h}_{ij} + \dot\varphi h_{ij} + \frac{1}{a^2}h_i{}^k\chi_{,jk} \bigg)
+ \frac{\chi^{,k}}{a^2} \Big(  h_{ik,j}  + h_{jk,i}  - h_{ij,k} \Big)
\bigg]
\nonumber\\
& + 3H \bigg[ 
  \dot{h}_{ij}   \alpha
+ 2 \bigg( \varphi\dot{h}_{ij} + \dot\varphi h_{ij} + \frac{1}{a^2}h_i{}^k\chi_{,jk} \bigg)
+ \frac{\chi^k}{a^2} \Big(  h_{ik,j}  + h_{jk,i}  - h_{ij,k} \Big)
\bigg]
\nonumber\\
& + \alpha \frac{d}{dt} \Big(  \dot{h}_{ij}  \Big)
- \frac{1}{a^2}\chi^{,k}   \dot{h}_{ij,k}  
+ \kappa   \dot{h}_{ij}  
+ \frac{1}{a^2} \bigg[
- 2   h_i{}^k  \alpha_{,jk}
- \Big(  h_{ik,j}  + h_{jk,i}  - h_{ij,k} \Big) \alpha^{,k}
\bigg]
\nonumber\\
& - \frac{\delta_{ij}}{3} \bigg[ \frac{d}{dt} \bigg(
   \frac{1}{a^2} 2h^{kl}\chi_{,kl} 
\bigg)
+ 3H \bigg(
   \frac{1}{a^2} 2h^{kl}\chi_{,kl} 
\bigg)
+ \frac{1}{a^2} \Big(
- 2   h^{kl}  \alpha_{,kl}
 \Big) \bigg]
+ \frac{1}{a^2}\chi_{,i}{}^{,k}   \dot{h}_{jk}  
- \frac{1}{a^2} \chi_{,j}{}^{,k}   \dot{h}_{ik}  
\nonumber\\
& 
+ \frac{1}{a^2} \bigg[
 2 \Big( -2\varphi\Delta{h}_{ij} + h_j{}^k\varphi_{,ik}
  - h_{ij}\Delta\varphi \Big)
+ \varphi^{,k} \Big( h_{ik,j} + h_{jk,i} - 3h_{ij,k} \Big)
- \frac{2}{3} 
h^{kl}\varphi_{,kl}\delta_{ij} - 2h^{kl}\varphi_{,kl}\delta_{ij}
\bigg]
\nonumber\\
& 
- 16\pi G \bigg[ \varphi\Pi_{ij}^{(t)} + \frac{1}{a^2} \bigg( h_i{}^k\Pi_{,ij} - \frac{1}{3}h_{ij}\Delta\Pi 
- \frac{\delta_{ij}}{3} h^{kl}\Pi_{,kl} \bigg) \bigg]
\, .
\end{align}

3. Tensor-tensor source $s_{ij}^{(tt)}$ denotes the collection of the products of two tensor perturbations:
\begin{align}
s_{ij}^{(tt)} =
& \frac{d}{dt} \Big( 
 2 h_i{}^k \dot{h}_{jk} 
\Big)
+ 3H \Big(
 2 h_i{}^k \dot{h}_{jk}  
\Big)
- \frac{\delta_{ij}}{3} \bigg[ \frac{d}{dt} \Big(
2 h^{kl} \dot{h}_{kl}
\Big)
+ 3H \Big(
 2 h^{kl} \dot{h}_{kl}
\Big)
\bigg]
\nonumber\\
& + \frac{1}{a^2} \bigg\{
2 h^{kl}
\Big( h_{il,jk}  + h_{jl,ik} 
 - h_{ij,kl}  - h_{kl,ij} \Big)
- 2   h_i{}^k    \Delta{h}_{jk} 
- h^{kl}{}_{,i} h_{kl,j} 
+ 2   h_i{}^{k,l}  
\Big(  h_{jl,k}  - h_{jk,l} \Big)
\nonumber\\
& \qquad
- \frac{\delta_{ij}}{3} \bigg[
-4   h^{kl}    \Delta{h}_{kl} 
+   h^{kl,m} 
\Big(   2h_{km,l}   - 3h_{kl,m} \Big)
\bigg]
\bigg\}
- 16\pi G \bigg( h_i{}^k\Pi_{jk}^{(t)} - \frac{\delta_{ij}}{3} h^{kl}\Pi_{kl}^{(t)} \bigg)
\, .
\end{align}

4. Scalar-vector source $s_{ij}^{(sv)}$ denotes the collection of the products of one of the scalar perturbations and vector perturbations:
\begin{align}
s_{ij}^{(sv)} =
& \frac{1}{a^3} \frac{d}{dt} \bigg\{ 
  a 
  \Big[ \chi^{(v)}_{(i,j)}  \alpha
+ 2\chi^{(v)}_{(i,j)}   \varphi
+   \varphi_{,j}\chi^{(v)}_i  + \varphi_{,i}\chi^{(v)}_j   \Big]
\bigg\}
\nonumber\\
& + \alpha \frac{d}{dt} \bigg(  \frac{1}{a^2}\chi^{(v)}_{(i,j)} \bigg)
- \frac{1}{a^4}  \chi^{,k}\chi^{(v)}_{(i,j)k} 
+   \frac{1}{a^2}\chi^{(v)}_{(i,j)} \kappa
 + \frac{1}{a^4} \Big(
2\chi^{(v)}_{k,(i}\chi_{,j)}{}^{,k}  + \chi^{,k}\chi^{(v)}_{k,ij}
\Big)
\nonumber\\
& - \frac{\delta_{ij}}{3} \bigg[ \frac{1}{a^3} \frac{d}{dt} \Big(
2a\chi^{(v)}_k\varphi^{,k} 
\Big)
 + \frac{1}{a^4} \Big(
2\chi^{(v)}_{k,l}\chi^{,kl} + \chi^{,k}\Delta\chi^{(v)}_k
 \Big) \bigg]
+ \frac{1}{a^4} \Big( \chi^{(v)}_{i,k}\chi_{,j}{}^{,k} + \chi_{,i}{}^{,k}\chi^{(v)}_{(j,k)} \Big) 
- \frac{1}{a^4} \Big( \chi^{(v)}_{k,j}\chi_{,i}{}^{,k} + \chi_{,j}{}^{,k}\chi^{(v)}_{(i,k)} \Big) 
\nonumber\\
& 
- 16\pi G \frac{\varphi}{a}\Pi_{(i,j)}^{(v)}
- 16 \pi G (\rho+p) \bigg( v_{,(i}v^{(v)}_{j)}  - \frac{\delta_{ij}}{3} v^{,k}v^{(v)}_k  \bigg)
\, .
\end{align}

5. Vector-vector source $s_{ij}^{(vv)}$ denotes the collection of the products of two vector perturbations:
\begin{align}
s_{ij}^{(vv)} =
& 
- \frac{1}{a^4} \chi^{(v)k} \chi^{(v)}_{(i,j)k} 
+ \frac{1}{a^4} \Big(   \chi^{(v)k}\chi^{(v)}_{k,ij} + \chi^{(v)k}{}_{,i}\chi^{(v)}_{k,j} \Big)
- \frac{\delta_{ij}}{3} \bigg[
 \frac{1}{a^4} 
 \Big( \chi^{(v)k,l}\chi^{(v)}_{k,l} + \chi^{(v)k}\Delta\chi^{(v)}_k \Big)
\bigg]
\nonumber\\
& 
+ \frac{1}{a^4}\chi^{(v)}_i{}^{,k}   \chi^{(v)}_{(j,k)} 
- \frac{1}{a^4} \chi^{(v)k}{}_{,j}    \chi^{(v)}_{(i,k)} 
+ 8 \pi G (\rho+p) \bigg[ v^{(v)}_iv^{(v)}_j - \frac{\delta_{ij}}{3} v^{(v)k}v^{(v)}_k \bigg]
\, .
\end{align}

6. Vector-tensor source $s_{ij}^{(vt)}$ denotes the collection of the products of one of the vector perturbations and tensor perturbations:
\begin{align}
s_{ij}^{(vt)} =
& \frac{d}{dt} \bigg\{ \frac{1}{a^2} \bigg[
 2h_i{}^k\chi^{(v)}_{(j,k)} 
+  \chi^{(v)k} \Big(  h_{ik,j}  + h_{jk,i}  - h_{ij,k} \Big)
\bigg]
\bigg\}
+ 3H \frac{1}{a^2} \bigg[
 2h_i{}^k\chi^{(v)}_{(j,k)} 
+  \chi^{(v)k} \Big(  h_{ik,j}  + h_{jk,i}  - h_{ij,k} \Big)
\bigg]
\nonumber\\
& 
- \frac{1}{a^2}\chi^{(v)k} \dot{h}_{ij,k}  
 - \frac{1}{a^2} \chi^{(v)k}{}_{,k}\dot{h}_{ij} 
- \frac{\delta_{ij}}{3} \bigg\{ \frac{d}{dt} \bigg[
 \frac{2}{a^2}   h^{kl}   \chi^{(v)}_{(k,l)} 
\bigg]
+ 3H \bigg[
 \frac{2}{a^2}   h^{kl}   \chi^{(v)}_{(k,l)} 
\bigg]
 \bigg\}
+ \frac{1}{a^2}\chi^{(v)}_i{}^{,k}   \dot{h}_{jk}  
- \frac{1}{a^2} \chi^{(v)k}{}_{,j}   \dot{h}_{ik}  
\nonumber\\
& 
- 16\pi G \frac{1}{a} \bigg( h_i{}^k\Pi_{(j,k)}^{(v)} - \frac{\delta_{ij}}{3} h^{kl}\Pi_{(k,l)}^{(v)} \bigg)
\, .
\end{align}

Having sorted out all possible second-order source terms, we can proceed to find the solution of the second-order induced GWs as follows. First, we solve the linear equations and obtain their solutions. Then these linear solutions can be used to obtain the explicit form of the sources. After the transverse-traceless projection of the source $s_{ij}$ [see \eqref{eq:source-proj}], we can solve the inhomogeneous equation for the tensor perturbations and obtain the analytic integral solutions.

\section{Linear solutions}
\label{sec:linearsol}

\subsection{Vector perturbations at linear order}

The following linear equations for the vector-type perturbations are derived respectively from the momentum constraint, traceless evolution equation \eqref{eq:traceless-eq2} and momentum conservation equation~\citep{2007PhRvD..76j3527H}:
\begin{align}
\frac{\Delta}{2a^3}\chi_i^{(v)} + 8\pi G(\rho+p)v_i^{(v)}
& =
0
\, ,
\\
\frac{1}{a^2} \frac{d}{dt} \Big( a\chi_i^{(v)} \Big) - 8\pi G\Pi_i^{(v)}
& =
0
\, ,
\\
\frac{1}{a^4(\rho+p)}\frac{d}{dt} \Big[ a^4(\rho+p)v_i^{(v)} \Big] + \frac{\Delta}{2a^2} \frac{\Pi_i^{(v)}}{\rho+p}
& = 
0
\, .
\end{align}
A great simplification is made in the case of vanishing vector-type stress, $\Pi_i^{(v)} = 0$. Then all the linear vector perturbations always vanish: 
\begin{equation}
\chi_i^{(v)} = v_i^{(v)} = 0 \, .
\end{equation}
So among the possible sources to the second-order GWs, scalar-vector, vector-vector and vector-tensor contributions are absent, and we have only scalar-scalar, scalar-tensor and tensor-tensor sources.

\subsection{Scalar perturbations at linear order}

With the perturbation in the extrinsic curvature $\kappa$ being written as
\begin{equation}
\label{eq:kappa}
\kappa = 3H\alpha - 3\dot\varphi - \frac{\Delta}{a^2}\chi \, ,
\end{equation}
the complete set of the linear equations for scalar perturbations is
\begin{align}
\label{eq:E-const}
4\pi G\delta\rho + H\kappa + \frac{\Delta}{a^2}\varphi 
& = 
0 
\, ,
\\
\label{eq:mom-const}
\kappa + \frac{\Delta}{a^2}\chi - 12\pi G (\rho+p) av 
& = 
0 
\, ,
\\
\label{eq:trace}
\dot\kappa + 2H\kappa - 4\pi G (\delta\rho + 3\delta{p}) + \bigg( 3\dot{H} + \frac{\Delta}{a^2} \bigg)\alpha 
& = 
0 
\, ,
\\
\label{eq:tracefree}
\dot\chi + H\chi - \varphi - \alpha - 8\pi G\Pi
& = 
0 
\, ,
\\
\label{eq:E-cons}
\dot{\delta\rho} + 3H(\delta\rho + \delta{p}) - (\rho+p) \bigg( \kappa - 3H\alpha + \frac{\Delta}{a}v \bigg) 
& = 
0 
\, ,
\\
\label{eq:mom-cons}
\frac{1}{a^4(\rho+p)} \frac{d}{dt} \Big[ a^4(\rho+p)v \Big] - \frac{1}{a}\alpha 
- \frac{1}{a(\rho+p)} \bigg( \delta{p} + \frac{2}{3}\frac{\Delta}{a^2}\Pi \bigg) 
& = 
0
\, .
\end{align}
While $\delta{p} = c_s^2\delta\rho + \tau\delta{S}$ with $\delta{S}$ being the entropy perturbation, for a barotropic fluid in the absence of $\delta{S}$, simply $c_s^2 = w$. Then, assuming no anisotropic stress ($\Pi = 0$) for simplicity, these equations become even simpler and allow analytic solutions.

\subsubsection{Solutions for linear scalar perturbations during MD}

The solutions of the scalar perturbations during MD are already given in \cite{1994ApJ...427..533H} up to linear order, in \cite{2012ApJ...752...50H} up to second order and in \cite{2016JCAP...07..017Y} up to third order respectively. The solutions can be written conveniently in terms of the curvature perturbation $\varphi$, which does not decay but remains constant during MD even on sub-horizon scales. Thanks to gauge transformations, the solutions in one gauge are enough to find those in other gauges.

We can readily solve the linear equations for the scalar perturbations during MD ($p=0$) without anisotropic stress ($\Pi = 0$). Summarizing, in the comoving gauge for which $v = \gamma = 0$, we find
\begin{equation}
\label{eq:com-scalarsol}
\varphi_v = \calR \, , 
\quad 
\alpha_v = 0 \, , 
\quad 
\chi_v = \frac{2}{5H} \calR \, , 
\quad
\kappa_v = \frac{2}{5} \frac{k^2}{a^2H} {\calR} \, , 
\quad 
\delta_v = \frac{2}{5} \frac{k^2}{a^2H^2} {\calR} \, .
\end{equation}
Here, the subscript $v$ means the solutions are written in the comoving gauge.
These solutions can be by gauge transformation used to obtain solutions in different gauge conditions, e.g. zero-shear gauge for which $\beta = \gamma = 0$ thus, as the name stands, $\chi = 0$ (thus a subscript $\chi$) and we find
\begin{equation}
\label{eq:zsg-scalarsol}
\varphi_\chi = \frac{3}{5}\calR \, , 
\quad 
\alpha_\chi = -\frac{3}{5}\calR \, , 
\quad 
\kappa_\chi = -\frac{9}{5}H\calR \, , 
\quad
v_\chi = -\frac{2}{5aH}\calR \, , 
\quad 
\delta_\chi = \frac{6}{5} \bigg( 1+\frac{k^2}{3a^2H^2} \bigg) {\calR} \, .
\end{equation}
Note that we can read easily the well-known relation during MD between the initial amplitude of the curvature perturbation in the comoving gauge $\calR$ and that in the zero-shear gauge, or the ``gravitational potential'' $\Phi = -\alpha_\chi = \varphi_\chi$, as
\begin{equation}
\Phi = \frac{3}{5}\calR \, .
\end{equation}

\subsubsection{Solutions for linear scalar perturbations during RD}

During RD, the linear equations of motion are most readily solvable in the zero-shear gauge. With $w=1/3$ and $a \propto \eta$ during RD, the equation of motion for the curvature perturbation $\varphi_\chi$ is obtained from the trace evolution equation combined with the energy constraint and traceless evolution equation as [see e.g. \cite{2005pfc..book.....M}]
\begin{equation}
\frac{d^2\varphi_\chi}{d\eta^2} + \frac{4}{\eta} \frac{d\varphi_\chi}{d\eta}
- \frac{\Delta}{3} \varphi_\chi = 0
\, .
\end{equation}
Then, with $z \equiv k\eta/\sqrt{3}$ where $1/\sqrt{3}$ is the sound speed during RD, we can straightforwardly find the linear solutions for the scalar perturbations:
\begin{equation}
\label{eq:scalarRD-ZSG}
\begin{split}
\varphi_\chi
& =
2\calR \frac{j_1(z)}{z}
\, ,
\\
v_\chi
& =
- \frac{1}{aH} \bigg[ j_0(z) - 2 \frac{j_1(z)}{z} \bigg] \calR
\, ,
\\
\kappa_\chi
& =
-6H \bigg[ j_0(z) - 2 \frac{j_1(z)}{z} \bigg] \calR
\, ,
\\
\alpha_\chi
& =
-\varphi_\chi
\, ,
\\
\delta_\chi
& =
4\calR \bigg[ - j_0(z) + 2\frac{j_1(z)}{z} + zj_1(z) \bigg]
\, ,
\end{split}
\end{equation}
where $j_n$ is the first-kind spherical Bessel function of order $n$. The solutions in other gauges can be obtained by appropriate gauge transformations, e.g. the curvature perturbation in the comoving gauge as $\varphi_v = \varphi - aHv$. In the comoving gauge, the linear solutions for the scalar perturbations are
\begin{equation}
\label{eq:scalarRD-com}
\begin{split}
\varphi_v
& =
\calR j_0(z)
\, ,
\\
\chi_v
& =
\frac{1}{H} \bigg[ j_0(z) - 2 \frac{j_1(z)}{z} \bigg] \calR
\, ,
\\
\kappa_v
& =
3H z^2 \bigg[ j_0(z) - 2 \frac{j_1(z)}{z} \bigg] \calR
\, ,
\\
\alpha_v
& =
\bigg[ 2j_0(z) - 4 \frac{j_1(z)}{z} - zj_1(z) \bigg] \calR
\, ,
\\
\delta_v
& =
4\calR \bigg[ - 2j_0(z) + 4\frac{j_1(z)}{z} + zj_1(z) \bigg]
\, .
\end{split}
\end{equation}
Here, we have set the coefficients in such a way that the initial amplitude of the curvature perturbation in the comoving gauge is, as for the solution during MD, $\calR$, i.e. $\lim_{z\to0} \varphi_v(z) = \calR$.
Note that since $\lim_{z\to0} j_0(z) = 1$ and $\lim_{z\to0} j_1(z)/z = 1/3$, we can find the well-known relation during RD between the comoving curvature perturbation $\calR$ and the gravitational potential $\Phi$ as $\Phi = 2\calR/3$.

\subsection{Tensor perturbations at linear order}

Decomposing the tensor perturbations in terms of the two polarization tensors $e_{ij}^+$ and $e_{ij}^\times$ in the Fourier space, 
\begin{equation}
h_{ij}(t,\bm{x}) = \int \frac{d^3k}{(2\pi)^3} e^{i\bm{k}\cdot\bm{x}} h_{ij}(t,\bm{k})
= \int \frac{d^3k}{(2\pi)^3} e^{i\bm{k}\cdot\bm{x}}
\Big[ h_+(t,\bm{k})e_{ij}^+(\bm{k}) + h_\times(t,\bm{k})e_{ij}^\times(\bm{k}) \Big] .
\end{equation}
Since the polarization tensors are orthogonal to each other, i.e.
\begin{equation}
e^{ij}_+ e_{ij}^\times = 0 
\quad \text{and} \quad
e^{ij}_+e_{ij}^+ = e^{ij}_\times e_{ij}^\times = 1
\, ,
\end{equation}
we can invert this to find
\begin{equation}
h_\lambda(t,\bm{k}) = e_\lambda^{ij}(\bm{k}) \int d^3x e^{-i\bm{k}\cdot\bm{x}} h_{ij}(t,\bm{x})
\end{equation}
for each polarization $\lambda$. Then the linear equation of motion for each polarization mode is identical as 
\begin{equation}
\ddot{h}+3H\dot{h} + \frac{k^2}{a^2}h = 0 
\, ,
\end{equation}
where we have omitted the polarization index $\lambda$. Introducing $v \equiv ah$ and moving to the conformal time $d\eta = dt/a$, the equation becomes
\begin{equation}
\label{eq:v-eq}
\frac{d^2v}{d\eta^2} + \bigg( k^2 - \frac{1}{a} \frac{d^2a}{d\eta^2} \bigg)v = 0 
\, .
\end{equation}

\subsubsection{Solutions for linear tensor perturbations during MD}

During MD, $a \propto \eta^2$ so that
\begin{equation}
\frac{1}{a}\frac{d^2a}{d\eta^2} = \frac{2}{\eta^2} \, .
\end{equation}
Thus in terms of a new variable $x \equiv k\eta$, \eqref{eq:v-eq} becomes
\begin{equation}
\frac{d^2v}{dx^2} + \bigg( 1 - \frac{2}{x^2} \bigg) v = 0 \, .
\end{equation}
The general solution of this equation is
\begin{equation}
\label{eq:v-MDsol}
v = c_1 xj_1(x) + c_2 xy_1(x)
\, ,
\end{equation}
where $y_n$ is the second-kind spherical Bessel function of order $n$. Since $\lim_{x\to0} y_1(x)/x \to -\infty$ and $\lim_{x\to0} j_1(x)/x = 1/3$, we choose $j_1(x)/x$ as the proper solution, with the value at $x\to0$ being the primordial value for the tensor perturbation $h^\lambda_0({\bm k})$ for each polarization $\lambda$: 
\begin{equation}
\label{eq:h-MDsol}
h_\lambda(\eta,{\bm k})
= 3h_0^\lambda({\bm k}) \frac{j_1(k\eta)}{k\eta}
\, .
\end{equation}

\subsubsection{Solutions for linear tensor perturbations during RD}

The basic equation of motion for the tensor perturbations is essentially the same as that during MD. That is, with $v \equiv ah$, we have the same linear equation for $v$ given by \eqref{eq:v-eq}. The only difference is that since $a \propto \eta$ during RD, with $x \equiv k\eta$, \eqref{eq:v-eq} is simply
\begin{equation}
\frac{d^2v}{dx^2} + v = 0 \, ,
\end{equation}
and the general solution is 
\begin{equation}
\label{eq:v-RDsol}
v = ah = c_1 xj_0(x) + c_2 xy_0(x) \, .
\end{equation}
Since $\lim_{x\to0} y_0(x) \to -\infty$ and $\lim_{x\to0} j_0(x) = 1$, we choose $j_0(x)$ as the proper solution, with the value at $x\to0$ being the primordial value for the tensor perturbation $h^\lambda_0({\bm k})$ for each polarization $\lambda$:
\begin{equation}
\label{eq:h-RDsol}
h_\lambda(\eta,{\bm k}) = h_0^\lambda({\bm k}) j_0(k\eta) \, .
\end{equation}

\section{Second-order solutions for induced GWs}
\label{sec:inducedGWsols}

\subsection{Equation of motion for tensor perturbations with source}

To extract only the tensor parts from \eqref{eq:traceless-eq2}, we apply the transverse-traceless projection so that $s_{ij}$ on the right-hand side becomes what only sources tensor perturbations, $s_{ij}^\text{(tensor)}$~\citep{2007PhRvD..76j3527H}:
\begin{align}
\label{eq:source-proj}
s_{ij}^\text{(tensor)} 
& \equiv 
s_{ij} - \frac{3}{2} \bigg( \partial_i\partial_j - \frac{\delta_{ij}}{3}\Delta \bigg) \Delta^{-2} \partial_k\partial_l s^{kl}
- 2 \Delta^{-1} \partial_{(i}\partial^ks_{j)k} + 2\Delta^{-2}\partial_i\partial_j \partial_k\partial_l s^{kl}
\nonumber\\
& = 
s_{ij} - 2 \Delta^{-1} \partial_{(i}\partial^k s_{j)k} 
+ \frac{1}{2}\Delta^{-2} \big( \partial_i\partial_j + \delta_{ij}\Delta \big) \partial_k\partial_l s^{kl} \, .
\end{align}
Since the two traceless polarization tensors are orthogonal to each other, we can extract the individual equation of each polarization mode $h_\lambda$ by multiplying the corresponding polarization tensor $e_{ij}^\lambda$. Moreover, since $e^{ij}_\lambda k_i = e^{ij}_\lambda k_j = 0$, we have in the Fourier space
\begin{align}
e^{ij}_\lambda(\bm{k}) \int d^3x e^{-i\bm{k}\cdot\bm{x}} s_{ij}^\text{(tensor)}(\bm{x})
& = 
e^{ij}_\lambda(\bm{k}) \int d^3x e^{-i\bm{k}\cdot\bm{x}} 
\int \frac{d^3q}{(2\pi)^3} e^{i\bm{q}\cdot\bm{x}} s_{ij}^\text{(tensor)}(\bm{q})
\nonumber\\
& = 
e^{ij}_\lambda(\bm{k}) \int d^3q 
\bigg[ s_{ij} - \frac{q_iq_k}{q^2} s^k{}_j  + \frac{q_jq_k}{q^2} s^k{}_i
+ \frac{1}{2} \bigg( -\frac{q_iq_jq_kq_l}{q^4} + \frac{q_kq_l}{q^2}\delta_{ij} \bigg) s^{kl} \bigg]
\int \frac{d^3x}{(2\pi)^3} e^{-i(\bm{k}-\bm{q})\cdot\bm{x}} 
\nonumber\\
& = 
e^{ij}_\lambda(\bm{k}) \, s_{ij}(\bm{k}) 
\, .
\end{align}
Thus, for each polarization $\lambda$ the equation of motion is
\begin{equation}
\label{eq:h-eom}
\ddot{h}_\lambda(t,\bm{k}) + 3H\dot{h}_\lambda(t,\bm{k}) + \frac{k^2}{a^2}h_\lambda(t,\bm{k})
= e^{ij}_\lambda(\bm{k}) \, s_{ij}(\bm{k}) \, .
\end{equation}

One more simplification is ahead. Since the source term $s_{ij}$ in the above equation is multiplied by the traceless polarization tensor $e^{ij}_\lambda$, the terms proportional to $\delta_{ij}$ in $s_{ij}$ identically vanish on the right-hand side of \eqref{eq:h-eom}. Thus, in the absence of the anisotropic stress, the source terms that survive \eqref{eq:h-eom} are
\begin{align}
\label{eq:nij-ss}
s_{ij}^{(ss)} 
= &
\frac{1}{a^3} \frac{d}{dt} \Big[
a \Big( 2\varphi\chi_{,ij} + \varphi_{,i}\chi_{,j} + \varphi_{,j}\chi_{,i} \Big)
\Big]
+ \frac{1}{a^2} \Big(
\kappa\chi_{,ij} - 4\varphi\varphi_{,ij}  -   3\varphi_{,i}\varphi_{,j}
\Big)
+ \frac{1}{a^4}\chi^{,k}{}_{,i}\chi_{,jk}
\nonumber\\
& 
+ \frac{1}{a^2}  \Big[ 2\alpha\dot\chi_{,ij} - H\alpha\chi_{,ij}  + \dot\alpha\chi_{,ij} 
- 2(\alpha+\varphi)\alpha_{,ij} - \alpha_{,i}\alpha_{,j} - 2\varphi_{(,i}\alpha_{,j)}
\Big]
+ 8\pi G (\rho+p) v_{,i}v_{,j} 
\, ,
\\
\label{eq:nij-tt}
s_{ij}^{(tt)} 
= & 
\frac{d}{dt} \Big( 
 2 h_i^k \dot{h}_{jk} 
\Big)
+ 3H \Big(
 2 h_i^k \dot{h}_{jk}  
\Big)
\nonumber\\
& + \frac{1}{a^2} \bigg[
2  h^{kl}
\Big(  h_{il,jk}  + h_{jl,ik} 
 - h_{ij,kl}  - h_{kl,ij} \Big)
- 2   h_i^k    \Delta{h}_{jk} 
- h^{kl}{}_{,i} h_{kl,j} 
+ 2   h_i^{k,l}  
\Big(  h_{jl,k}  - h_{jk,l} \Big) \bigg]
\, ,
\\
\label{eq:nij-st}
s_{ij}^{(st)} 
= & 
\frac{d}{dt} \bigg[ 
 \dot{h}_{ij}   \alpha
+ 2 \bigg( \varphi\dot{h}_{ij} + \dot\varphi h_{ij} + \frac{1}{a^2}h_i^k\chi_{,jk} \bigg)
+ \frac{\chi^{,k}}{a^2} \Big(  h_{ik,j}  + h_{jk,i}  - h_{ij,k} \Big)
\bigg]
\nonumber\\
& + 3H \bigg[ 
  \dot{h}_{ij}   \alpha
+ 2 \bigg( \varphi\dot{h}_{ij} + \dot\varphi h_{ij} + \frac{1}{a^2}h_i^k\chi_{,jk} \bigg)
+ \frac{\chi^{,k}}{a^2} \Big(  h_{ik,j}  + h_{jk,i}  - h_{ij,k} \Big)
\bigg]
\nonumber\\
& + \alpha \frac{d}{dt} \bigg(  \dot{h}_{ij}  \bigg)
- \frac{1}{a^2}\chi^{,k}   \dot{h}_{ij,k}  
+ \kappa   \dot{h}_{ij}  
+ \frac{1}{a^2} \bigg[
- 2   h_i^k  \alpha_{,jk}
- \Big(  h_{ik,j}  + h_{jk,i}  - h_{ij,k} \Big) \alpha^{,k}
\bigg]
+ \frac{1}{a^2}\chi_{,i}{}^{,k}   \dot{h}_{jk}  
- \frac{1}{a^2} \chi_{,j}{}^{,k}   \dot{h}_{ik}  
\nonumber\\
& 
+ \frac{1}{a^2} \bigg[
 2 \Big( -2\varphi\Delta{h}_{ij} + h_j^k\varphi_{,ik}
  - h_{ij}\Delta\varphi \Big)
+ \varphi^{,k} \Big( h_{ik,j} + h_{jk,i} - 3h_{ij,k} \Big)
\bigg]
\, .
\end{align}

To find the solution for this inhomogeneous equation, we use the Green's function solution [for earlier attempts to apply the Green's function solutions to cosmological perturbations, see~\cite{2001PhLB..510....1S,2002PhLB..538..213G}]. That is, let $L$ be a linear second-order differential operator, and the equation we want to solve is of the form
\begin{equation}
Ly(x) = r(x) \, ,
\end{equation}
with the two homogeneous solutions being $y_1$ and $y_2$. Then, the full solution is given by
\begin{align}
y(x) 
& =  
(\text{appropriate combination of $y_1$ and $y_2$ according to the boundary conditions})
\nonumber\\
& 
\quad
+ \int d\tilde{x} r(\tilde{x}) \underbrace{ 
\frac{y_1(\tilde{x})y_2(x) - y_2(\tilde{x})y_1(x)}{y_1(\tilde{x})y_2'(\tilde{x}) - y_2(\tilde{x})y_1'(\tilde{x})}
}_{\equiv G(x,\tilde{x})} 
\, .
\end{align}
We have seen that the homogeneous solutions during MD are given by \eqref{eq:v-MDsol}, then the Green's function during MD becomes
\begin{equation}
\label{eq:Green-MD}
G_\text{MD}(\eta,\tilde\eta) 
= 
\frac{x\tilde{x}}{k} \Big[ j_1(\tilde{x})y_1(x) - j_1(x)y_1(\tilde{x}) \Big] 
\, .
\end{equation}
With the homogeneous solutions during RD being given by \eqref{eq:v-RDsol}, the Green's function during RD becomes
\begin{equation}
\label{eq:Green-RD}
G_\text{RD}(\eta,\tilde\eta)
=
\frac{x\tilde{x}}{k} \Big[ j_0(\tilde{x})y_0(x) - j_0(x)y_0(\tilde{x}) \Big]
\, .
\end{equation}
Comparing with \eqref{eq:Green-MD}, the only difference is the order of the spherical Bessel functions inside the square brackets.

Thus, we expect the second-order induced GWs sourced by the product of two linear perturbations $X$ and $Y$ would be, during both MD and RD, of the general integral form:
\begin{align}
\label{eq:generalsol}
h_\lambda(\eta,{\bm k})
& =
\frac{1}{a} \int_0^\eta d\tilde\eta \, \Big[ a^3(\tilde\eta)  e^{ij}_\lambda(\bm{k}) s_{ij}(\bm{k}) \Big]
G(\eta,\tilde\eta)
\nonumber\\
& =
\int \frac{d^3q}{(2\pi)^3} \Big[ e^{ij}_\lambda(\bm{k}) (\cdots)_{ij} \Big]
X_0({\bm k}-{\bm q}) Y_0({\bm q}) 
\int_0^x d\tilde{x} K(\tilde{x},{\bm k},{\bm q})
\, .
\end{align}
Here, the integral over an internal momentum ${\bm q}$ is because the source $s_{ij}$ is the product of two perturbations $X$ and $Y$, it is written as a convolution in the Fourier space. The terms inside the square brackets constitute the (dimensionless) projection of the polarization tensor $e^{ij}_\lambda(\bm{k})$. $X_0({\bm k})$ and $Y_0({\bm k})$ denote respectively the initial amplitudes of $X$ and $Y$, i.e. $\calR({\bm k})$ and/or $h_0^\lambda({\bm k})$. Finally, the integral over $\tilde{x}$ is the kernel which is a function of both momenta as well as time. Our main concern in finding this analytic integral solution is to compute this kernel. In the following, we proceed with the sources \eqref{eq:nij-ss}, \eqref{eq:nij-tt} and \eqref{eq:nij-st} to calculate the closed analytic form of the kernel.

\subsection{Scalar-scalar induced GWs during MD}

We first consider the scalar-scalar source \eqref{eq:nij-ss}. The analytic integral solutions in various gauge conditions are given only very recently in~\cite{2017ApJ...842...46H} so we can check our results in this section. We consider only two gauges, comoving and zero-shear gauges in which the solutions of the linear scalar perturbations during MD are given respectively by \eqref{eq:com-scalarsol} and \eqref{eq:zsg-scalarsol}. We first compute the Fourier component of $s_{ij}^{(ss)}$ in the zero-shear gauge for which $\chi = 0$ and $\alpha = -\varphi$, so $s_{ij}^{(ss)}$ is greatly simplified. After straightforward calculations we find in the zero-shear gauge the source $s_{ij}^{(ss)}$ purely in terms of the initial perturbation $\calR$ as
\begin{equation}
\label{eq:zsg-ss-source}
e_\lambda^{ij}(\bm{k}) s_{ij}^{(ss)}(\bm{k})
= 
\frac{1}{a^2} \frac{6}{5} \int \frac{d^3q}{(2\pi)^3} \big[ e_\lambda^{ij}(\bm{k}) q_iq_j \big] 
\calR(\bm{k}-\bm{q})\calR(\bm{q})
\, .
\end{equation}
Note that other than the overall $1/a^2$, there is no time dependence. 
Then for each polarization, the solution of the GWs induced by the scalar-scalar source in the zero-shear gauge is
\begin{align}
\label{eq:h2nd-zsgsol}
h_\lambda(\eta,\bm{k}) 
& = 
\frac{6}{5} \bigg[ 1 - 3\frac{j_1(k\eta)}{k\eta} \bigg] 
\int \frac{d^3q}{(2\pi)^3} \left[ e_\lambda^{ij}(\bm{k}) \frac{q_iq_j}{k^2} \right] 
\calR(\bm{k}-\bm{q})\calR(\bm{q})
\, .
\end{align}
Compared with the general form of the solution \eqref{eq:generalsol}, the kernel is a function of $k$ and $\eta$ in the specific combination $k\eta$, and thus can be pulled out of the internal momentum integral.

\begin{figure}[ht!]
\begin{center}
\includegraphics[width=0.45\textwidth]{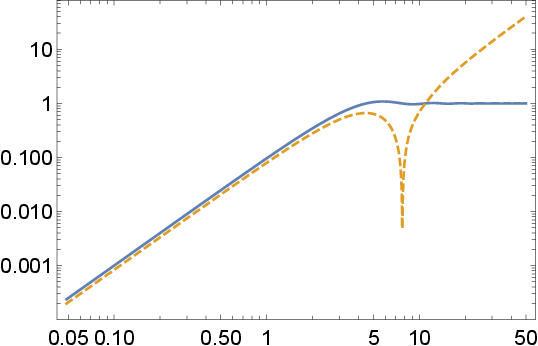}
\end{center}
\caption{The kernels for the scalar-scalar induced GWs during MD in the zero-shear gauge \eqref{eq:h2nd-zsgsol} (solid line) and comoving gauge \eqref{eq:h2nd-comsol} (dashed line) as a function of $k\eta$.}
\label{fig:MDss}
\end{figure}

Likewise, in the comoving gauge, we have $v=0$ and $\alpha=0$ during MD, so that the scalar-scalar source \eqref{eq:nij-ss} becomes
\begin{equation}
e^{ij}_\lambda(\bm{k}) s_{ij}^{(ss)} 
= 
\frac{1}{a^2} \int \frac{d^3q}{(2\pi)^3} \big[ e^{ij}_\lambda(\bm{k}) q_iq_j \big]
\bigg( 1 - \frac{2}{25} \frac{k^2}{a^2H^2} \bigg)
\calR(\bm{k}-\bm{q})\calR(\bm{q})
\, ,
\end{equation}
where the additional term comes from the spatial gradient of the shear. Then we can find trivially the solution as, using $aH=2/\eta$ during MD,
\begin{align}
\label{eq:h2nd-comsol}
h_\lambda(\eta,\bm{k}) 
& = 
\frac{6}{5} \bigg[ 1 - 3\frac{j_1(k\eta)}{k\eta}  - \frac{(k\eta)^2}{60} \bigg]
\int \frac{d^3q}{(2\pi)^3} \left[ e_\lambda^{ij}(\bm{k}) \frac{q_iq_j}{k^2} \right] 
\calR(\bm{k}-\bm{q})\calR(\bm{q})
\, .
\end{align}
Again, the kernel is a function of only $k\eta$. The reason why we have such a simple kernel for the scalar-scalar source is because the scalar perturbations can be written in terms of the constant $\calR$, so the time integral is greatly simplified. These scalar-scalar induced solutions in the zero-shear gauge \eqref{eq:h2nd-zsgsol} and comoving gauge \eqref{eq:h2nd-comsol} agree with~\cite{2017ApJ...842...46H}. In Figure~\ref{fig:MDss}, we show the kernels barring the factor 6/5. As $k\eta$ becomes bigger, the zero-shear gauge kernel approaches 1 while that in the comoving gauge increases as $(k\eta)^2$. Thus on small scales the amplitude of the induced GWs in the comoving gauge is much bigger than that in the zero-shear gauge. This shows clearly the gauge dependence of the scalar-induced GWs.

\subsection{Scalar-tensor induced GWs during MD}

Next we consider the GWs induced by the scalar-tensor source during MD.
Considering first the zero-shear gauge in which $\chi = 0$ and $\alpha = - \varphi$, we find (omitting the subscript $\chi$ for the scalar perturbation $\varphi$)
\begin{align}
s_{ij}^{(st)}({\bm k}) 
& =
\int \frac{d^3q_1d^3q_2}{(2\pi)^3} \delta^{(3)}({\bm k}-{\bm q}_{12})
\frac{1}{2} \bigg[
\bigg( \frac{d}{dt} + 3H \bigg) \Big( \varphi_1 \dot{h}_{\lambda'2} 
+ 2\dot\varphi_1 h_{\lambda'2} \Big)
- \varphi_1 \frac{d}{dt} \Big( \dot{h}_{\lambda'2} \Big)
+ \kappa_1 \dot{h}_{\lambda'2} 
+ \frac{2}{a^2} \Big( 2 k^k q_{2k} + q_1^2 \Big) \varphi_1 h_{\lambda'2}
\bigg] e_{ij2}^{\lambda'}
\nonumber\\
& \quad
+ ({\bm q}_1 \leftrightarrow {\bm q}_2) 
\, ,
\end{align}
where the subscript 1 for a perturbation variable means it is a function of ${\bm q}_1$, e.g. $\varphi_1 = \varphi({\bm q}_1)$ and so on.
Using $\varphi_\chi = 3\calR/5$ and $\kappa_\chi = -9H\calR/5$ during MD with $\calR$ = constant, the terms including the time derivative of the linear tensor perturbation vanish and we have the following simple expression for the source:
\begin{equation}
e^{ij}_\lambda({\bm k})
s_{ij}^{(st)}({\bm k}) 
=
\frac{1}{a^2} \frac{6}{5}
\int \frac{d^3q}{(2\pi)^3} 
\big[ e^{ij}_\lambda({\bm k}) e_{ij}^{\lambda'}({\bm q}) \big]
\big( k^2 + q^2 \big) \calR({\bm k}-{\bm q}) h_{\lambda'}(\eta,{\bm q}) 
\, .
\end{equation}
It is very important to note that the linear order perturbation $h_{\lambda'}$ does possess time dependence as given by \eqref{eq:h-MDsol}. Thus, unlike the scalar-scalar source case, the time integral for the Green's function solution includes another time-dependent function from $h_{\lambda'}$, namely, $j_1(k\eta)/(k\eta)$:
\begin{align}
\label{eq:h2nd-STsolZSG-MD}
\hspace{-.5em}
h_\lambda(\eta,{\bm k})
& =
\frac{18}{5} \int \frac{d^3q}{(2\pi)^3} 
\big[ e^{ij}_\lambda({\bm k}) e_{ij}^{\lambda'}({\bm q}) \big]
\big( q^2 + k^2 \big) \calR({\bm k}-{\bm q}) h_0^{\lambda'}({\bm q}) 
\frac{1}{kqx} \bigg[
y_1(x) \int_0^x d\tilde{x} \tilde{x}^2 j_1\bigg( \frac{q}{k}\tilde{x} \bigg) j_1(\tilde{x})
- j_1(x) \int_0^x d\tilde{x} \tilde{x}^2 j_1\bigg( \frac{q}{k}\tilde{x} \bigg) y_1(\tilde{x})
\bigg]
\nonumber\\
& = 
\frac{18}{5} \int \frac{d^3q}{(2\pi)^3} 
\big[ e^{ij}_\lambda({\bm k}) e_{ij}^{\lambda'}({\bm q}) \big]
\calR({\bm k}-{\bm q}) h_0^{\lambda'}({\bm q}) 
\bigg[  \frac{j_1(k\eta)}{k\eta} - \frac{j_1(q\eta)}{q\eta} \bigg]
\frac{q^2+k^2}{q^2-k^2}  
\, .
\end{align}
The detail of the $\tilde{x}$-integrals is given in Appendix~\ref{app:MDss-st}.

\begin{figure}[ht!]
\begin{center}
\includegraphics[width=0.45\textwidth]{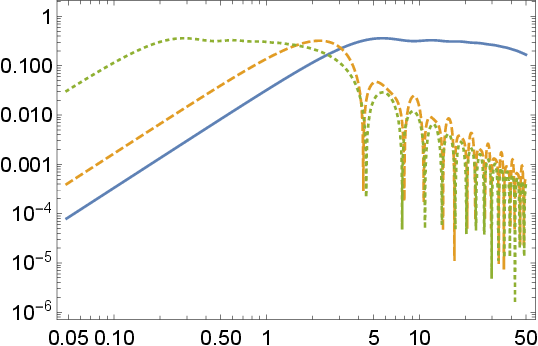}
\hspace{2em}
\includegraphics[width=0.45\textwidth]{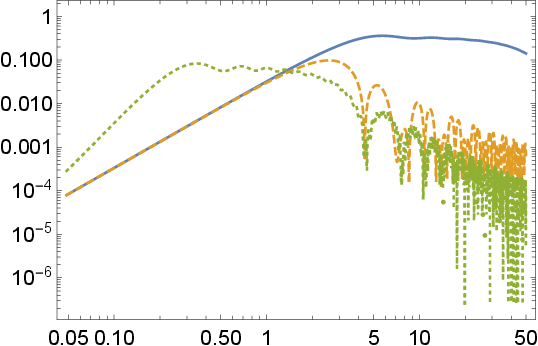}
\end{center}
\caption{The kernels in the (left) zero-shear gauge and (right) comoving gauge as a function of $k\eta$. Since the kernels are also dependent on $q$, we set (solid lines) $q/k = 0.05$, (dashed lines) $q/k = 2$ and (dotted lines) $q/k = 20$ in both panels.}
\label{fig:MDst}
\end{figure}

In the comoving gauge where $v=0$, using $\calR=$ constant and 
\begin{equation}
\frac{d}{dt} \bigg( \frac{1}{a^2H} \bigg)
=
-\frac{1}{2a^2} 
\, ,
\end{equation}
which follows from $H = 2/(3t)$ during MD, straightforward calculations give
\begin{equation}
e^{ij}_\lambda({\bm k}) s_{ij}^{(st)}({\bm k})
=
\frac{2}{a^2} \int \frac{d^3q}{(2\pi)^3} 
\big[ e^{ij}_\lambda({\bm k}) e^{\lambda'}_{ij}({\bm q}) \big]
\calR({\bm k}-{\bm q})
\bigg[
\frac{1}{5H} \big( k^2 - q^2 \big) \dot{h}_{\lambda'}({\bm q}) 
+ k^2 {h}_{\lambda'}({\bm q}) 
\bigg]
\, .
\end{equation}
Unlike the zero-shear gauge case, the time derivative of the linear tensor perturbation remains, which is given by
\begin{align}
\label{eq:h-deriv}
\dot{h}_{\lambda'}({\bm q})
=
\frac{q}{a} \frac{d}{d(q\eta)} h_{\lambda'}(\eta,{\bm q})
=
\frac{q}{a} 3 h_0^{\lambda'}({\bm q}) 
\bigg[ -3\frac{j_1(q\eta)}{(q\eta)^2} + \frac{j_0(q\eta)}{q\eta} \bigg]
\, .
\end{align}
We can find the solution in a manner very similar to \eqref{eq:h2nd-STsolZSG-MD} but with different $s_{ij}^{(st)}$ as:
\begin{align}
\label{eq:h2nd-STsolCG-MD}
h_\lambda(\eta,{\bm k})
& =
\frac{6}{k^2x} 
\int \frac{d^3q}{(2\pi)^3} 
\big[ e^{ij}_\lambda({\bm k}) e_{ij}^{\lambda'}({\bm q}) \big]
\calR({\bm k}-{\bm q}) h_0^{\lambda'}({\bm q})
\int_0^x d\tilde{x} 
\tilde{x}^3
\bigg[ \frac{3q^2+7k^2}{10} \frac{j_1(q\tilde{x}/k)}{q\tilde{x}/k}
+ \frac{k^2-q^2}{10} j_0\bigg( \frac{q}{k} \tilde{x} \bigg) \bigg]
\nonumber\\
& \quad \times
\Big[ j_1(\tilde{x})y_1(x) - j_1(x)y_1(\tilde{x}) \Big]
\nonumber\\
& =
6
\int \frac{d^3q}{(2\pi)^3} 
\big[ e^{ij}_\lambda({\bm k}) e_{ij}^{\lambda'}({\bm q}) \big]
\calR({\bm k}-{\bm q}) h_0^{\lambda'}({\bm q})
\bigg[
\frac{q^2+5k^2}{5(q^2-k^2)} \frac{j_1(k\eta)}{k\eta} 
- \frac{5q^2+7k^2}{10(q^2-k^2)} \frac{j_1(q\eta)}{q\eta} + \frac{1}{10} j_0(q\eta) 
\bigg]
\, .
\end{align}
This is the solution of the induced $h_\lambda({\bm k},\eta)$ from the scalar-tensor source in the comoving gauge. In Figure~\ref{fig:MDst} we show the kernels in both gauges. For $k\eta \gtrsim 1$ with sizeable $q/k$, the comoving gauge kernel exhibits more rapid oscillations. But for small $q/k$, both kernels are approximated by $1/3-j_1(k\eta)/(k\eta) + \calO(q^2/k^2)$ so they behave similarly. Compared to the scalar-scalar induced GWs as we have seen previously, the difference in the zero-shear gauge and comoving gauge is not prominent. This is because the gradient of the shear, that gives rise to the huge difference on small scales for the scalar-scalar induced GWs, is highly suppressed by the exponentially decaying linear GWs on small scales.

\subsection{Tensor-tensor induced GWs during MD}

Next we consider the tensor-tensor source during MD. Multiplying the polarization tensor $e^{ij}_\lambda({\bm k})$ and separating the time-dependent part gives
\begin{align}
\hspace{-3em}
e^{ij}_\lambda({\bm k}) s_{ij}^{(tt)}({\bm k})
& =
\frac{1}{a^2} \int \frac{d^3q_1d^3q_2}{(2\pi)^3} \delta^{(3)}({\bm k}-{\bm q}_{12})
\bigg\{ 
\underbrace{ 
9 h_0^{\lambda_1}({\bm q}_1) h_0^{\lambda_2}({\bm q}_2)
e^{ij}_\lambda({\bm k}) e_{i1}^{k\lambda_1} e_{jk2}^{\lambda_2} 
}_{\equiv \calC_\lambda({\bm q}_1,{\bm q}_2) } q_1q_2 
\bigg[ -3\frac{j_1(q_1\eta)}{(q_1\eta)^2} + \frac{j_0(q_1\eta)}{q_1\eta} \bigg]
\bigg[ -3\frac{j_1(q_2\eta)}{(q_2\eta)^2} + \frac{j_0(q_2\eta)}{q_2\eta} \bigg]
\nonumber\\
&
\hspace{10em}
+ 
9 h_0^{\lambda_1}({\bm q}_1) h_0^{\lambda_2}({\bm q}_2) e^{ij}_\lambda({\bm k})
\bigg[
e^{kl}_{\lambda_11} \big(
-q_{2j}q_{2k} e^{\lambda_2}_{il2} - q_{2i}q_{2k} e_{jl2}^{\lambda_2} 
+ q_{2k}q_{2l} e^{\lambda_2}_{ij2} 
\big)
\nonumber\\
&
\hspace{11em}
\underbrace{ 
\hspace{10em}
+ \frac{1}{2} q_{1i}q_{1j} e_{\lambda_1}^{kl1} e_{kl2}^{\lambda_2} 
+ ({\bm q}_1\cdot{\bm q}_2) e_{i1}^{k\lambda_1} e_{jk2}^{\lambda_2}
- q_1^lq_{2k} e_{i1}^{k\lambda_1} e_{jl2}^{\lambda_2} \bigg] 
}_{\equiv \calD_\lambda({\bm q}_1,{\bm q}_2) }
\frac{j_1(q_1\eta)}{q_1\eta} \frac{j_1(q_2\eta)}{q_2\eta} 
\bigg\} 
\nonumber\\
& 
\quad
+ ({\bm q}_1 \leftrightarrow {\bm q}_2) 
\, .
\end{align}
As can be seen, only the Bessel function terms contain time dependence. Then the solution for $h_\lambda$ can be written as
\begin{align}
\label{eq:tt-sol1}
h_\lambda(\eta,{\bm k}) 
& =
\int \frac{d^3q_1d^3q_2}{(2\pi)^3} \delta^{(3)}({\bm k}-{\bm q}_{12})
\Bigg\{
\Big[ \calC_\lambda({\bm q}_1,{\bm q}_2) + ({\bm q}_1 \leftrightarrow {\bm q}_2) \Big]
\frac{\bar{q}_1\bar{q}_2}{x} \bigg[
\frac{9}{\bar{q}_1^2\bar{q}_2^2} \int_0^x d\tilde{x} \tilde{x}^{-1} j_1(\bar{q}_1\tilde{x}) j_1(\bar{q}_1\tilde{x})
- \frac{3}{\bar{q}_1^2\bar{q}_2} \int_0^x d\tilde{x} j_1(\bar{q}_1\tilde{x}) j_0(\bar{q}_2\tilde{x})
\nonumber\\
& 
\hspace{14em}
- \frac{3}{\bar{q}_1\bar{q}_2^2} \int_0^x d\tilde{x} j_0(\bar{q}_1\tilde{x}) j_1(\bar{q}_2\tilde{x})
+ \frac{1}{\bar{q}_1\bar{q}_2} \int_0^x d\tilde{x} \tilde{x} j_0(\bar{q}_1\tilde{x}) j_0(\bar{q}_2\tilde{x})
\bigg]
\Big[ j_1(\tilde{x})y_1(x) - j_1(x)y_1(\tilde{x}) \Big] 
\nonumber\\
& 
\hspace{10em}
+
\Big[ \calD_\lambda({\bm q}_1,{\bm q}_2) + ({\bm q}_1 \leftrightarrow {\bm q}_2) \Big]
\frac{1}{k^2x} \frac{1}{\bar{q}_1\bar{q}_2} \int_0^x d\tilde{x} \tilde{x} j_1(\bar{q}_1\tilde{x}) j_1(\bar{q}_1\tilde{x})
\Big[ j_1(\tilde{x})y_1(x) - j_1(x)y_1(\tilde{x}) \Big] 
\Bigg\}
\, ,
\end{align}
where $\bar{q}_1 \equiv q_1/k$ and $\bar{q}_2 \equiv q_2/k$ respectively.

\begin{figure}[ht!]
\begin{center}
\includegraphics[width=0.45\textwidth]{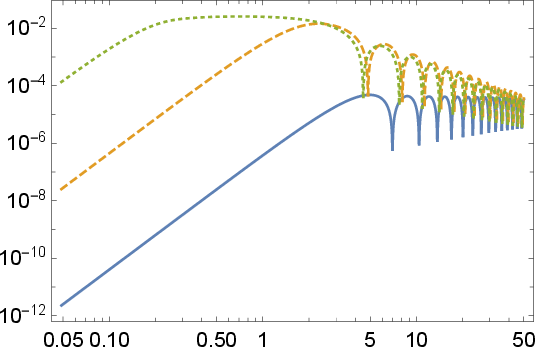}
\hspace{2em}
\includegraphics[width=0.45\textwidth]{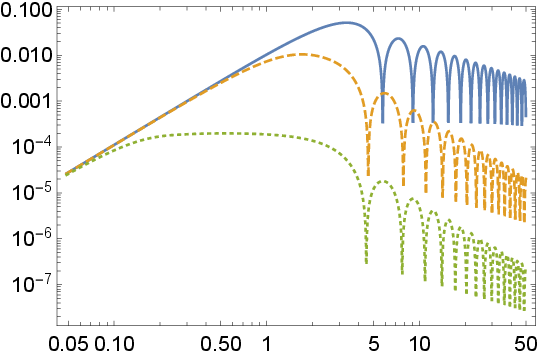}
\end{center}
\caption{The kernels (left) $F_\text{MD}$ and (right) $G_\text{MD}$ as a function of $k\eta$. Since they are also dependent on the angle between ${\bm k}$ and ${\bm q}$, for simplicity we set in such a way that for both (solid lines) $q/k = 0.05$ and (dotted lines) $q/k = 20$ they are aligned perpendicularly, i.e. $\cos\big(\hat{\bm k}\cdot\hat{\bm q}\big) = 0$, while for (dashed lines) $q/k = 2$ the angle between them is $2\pi/3$, $\cos\big(\hat{\bm k}\cdot\hat{\bm q}\big) = -1/2$ in both panels.}
\label{fig:MDtt}
\end{figure}

Now, using the recurrence relation \eqref{eq:recurrence} for $n=1$, we find
\begin{equation}
j_2(\bar{q}_2x) 
= 
\frac{3}{\bar{q}_2x} j_1(\bar{q}_1x) - j_0(\bar{q}_1x) 
= 
\frac{3}{\bar{q}_2} \bigg[ \frac{j_1(\bar{q}_2x)}{x} - \frac{\bar{q}_2}{3}j_0(\bar{q}_2x) \bigg]
\, ,
\end{equation}
so that the terms multiplied by $\calC_\lambda({\bm q}_1,{\bm q}_2)$ become very simple as
\begin{align}
& 
\bigg[
\frac{9}{\bar{q}_1^2\bar{q}_2^2} \int_0^x d\tilde{x} \tilde{x}^{-1} j_1(\bar{q}_1\tilde{x}) j_1(\bar{q}_2\tilde{x})
- \frac{3}{\bar{q}_1^2\bar{q}_2} \int_0^x d\tilde{x} j_1(\bar{q}_1\tilde{x}) j_0(\bar{q}_2\tilde{x})
- \frac{3}{\bar{q}_1\bar{q}_2^2} \int_0^x d\tilde{x} j_0(\bar{q}_1\tilde{x}) j_1(\bar{q}_2\tilde{x})
+ \frac{1}{\bar{q}_1\bar{q}_2} \int_0^x d\tilde{x} \tilde{x} j_0(\bar{q}_1\tilde{x}) j_0(\bar{q}_2\tilde{x})
\bigg]
f_1(\tilde{x})
\nonumber\\
& =
\frac{1}{\bar{q}_1\bar{q}_2} \int_0^x d\tilde{x} \tilde{x} j_2(\bar{q}_1\tilde{x}) j_2(\bar{q}_2\tilde{x}) f_1(\tilde{x})
\, ,
\end{align}
where $f_1(\tilde{x})$ denotes both first- and second-kind of the spherical Bessel functions. Since $\lim_{x\to0}y_1(x) = -1/x^2-1/2+\cdots$ and $\lim_{x\to0}j_n(x) \sim x^n$, we always have converging results. Thus, we can write \eqref{eq:tt-sol1} as
\begin{align}
\label{eq:tt-source-sol}
h_\lambda(\eta,{\bm k}) 
& =
\int \frac{d^3q_1d^3q_2}{(2\pi)^3} \delta^{(3)}({\bm k}-{\bm q}_{12})
\bigg\{
\Big[ \calC_\lambda({\bm q}_1,{\bm q}_2) + ({\bm q}_1 \leftrightarrow {\bm q}_2) \Big]
\frac{1}{x} \int_0^x d\tilde{x} \tilde{x} j_2(\bar{q}_1\tilde{x}) j_2(\bar{q}_2\tilde{x})
\Big[ j_1(\tilde{x})y_1(x) - j_1(x)y_1(\tilde{x}) \Big] 
\nonumber\\
& 
\hspace{12em}
+
\frac{1}{k^2} \Big[ \calD_\lambda({\bm q}_1,{\bm q}_2) + ({\bm q}_1 \leftrightarrow {\bm q}_2) \Big]
\frac{1}{\bar{q}_1\bar{q}_2x} \int_0^x d\tilde{x} \tilde{x} j_1(\bar{q}_1\tilde{x}) j_1(\bar{q}_2\tilde{x})
\Big[ j_1(\tilde{x})y_1(x) - j_1(x)y_1(\tilde{x}) \Big] 
\bigg\}
\, .
\end{align}
Both the $\tilde{x}$ integrals can be performed analytically, with the details given in Appendix~\ref{app:integrals-MDtt}. Performing the momentum integral using the delta function, finally \eqref{eq:tt-source-sol} becomes
\begin{align}
\label{eq:ttsol-explicit2}
h_\lambda(\eta,{\bm k}) 
& =
\int \frac{d^3q}{(2\pi)^3} 18 h_0^{\lambda_1}({\bm q}) h_0^{\lambda_2}({\bm k}-{\bm q})
\Bigg\{
e_\lambda^{ij}({\bm k}) e_i^{k\lambda_1}({\bm q}) e_{jk}^{\lambda_2}({\bm k}-{\bm q}) 
F_\text{MD}({\bm k},{\bm q},\eta)
\nonumber\\
& 
\hspace{1.5em}
+
\frac{1}{k^2} 
\bigg(
e_\lambda^{ij}({\bm k})
e^{kl}_{\lambda_1}({\bm q}) \Big(
q_jk_k e_{il}^{\lambda_2}({\bm k}-{\bm q}) 
+ q_ik_k e_{jl}^{\lambda_2}({\bm k}-{\bm q})
+ k_kk_l e_{ij}^{\lambda_2}({\bm k}-{\bm q})
\Big)
\nonumber\\
&
\hspace{2.em}
+
e_\lambda^{ij}({\bm k}) 
\bigg[
\frac{1}{2} q_i q_j e^{kl}_{\lambda_1}({\bm q}) e_{kl}^{\lambda_2}({\bm k}-{\bm q})
+ {\bm q}\cdot({\bm k}-{\bm q}) e_i^{k\lambda_1}({\bm q}) e_{jk}^{\lambda_2}({\bm k}-{\bm q})
- q^lk_k e_i^{k\lambda_1}({\bm q}) e_{jl}^{\lambda_2}({\bm k}-{\bm q})
\bigg]
\bigg) 
G_\text{MD}({\bm k},{\bm q},\eta)
\Bigg\}
\, ,
\end{align}
where the kernels $F_\text{MD}$ and $G_\text{MD}$ are given respectively by \eqref{eq:j2j2-integral} and \eqref{eq:j1j1-integral}. In Figure~\ref{fig:MDtt} we show $F_\text{MD}$ and $G_\text{MD}$.

\subsection{Scalar-scalar induced GWs during RD}

Until now we have considered the second-order solutions for the induced GWs during MD. Now we consider the solutions during RD. Especially, after the first version of the present work appeared in arXiv, there have been a number of papers published in the literature, e.g.~\cite{2020JCAP...03..014D,2020PhRvD.101b3523I,2020PhRvD.101f3018Y}, claiming that the gauge dependence of the induced GWs during RD disappears. This is not the case, at least regarding the solutions in the two gauge conditions we examine here explicitly -- the zero-shear gauge and comoving gauge: the solutions are clearly different in these two gauges as can be seen in \eqref{eq:h2nd-RHS} and \eqref{eq:h2nd-ssCG} [see also~\cite{2020PhRvD.101h3529T}].

An important difference for the linear solutions for the scalar perturbations is that now the curvature perturbation $\varphi$ does not stay constant, but decays once the mode enters the horizon. Thus, the scalar-scalar induced GWs during RD do not behave simply as we have seen during MD, but exhibit rapid oscillations. 
We first consider the zero-shear gauge, in which the scalar-scalar source reads rather simply as
\begin{equation}
s_{ij}^{(ss)}(\bm{k}) 
= 
\int \frac{d^3q}{(2\pi)^3}
\bigg[
\frac{2}{a^2} q_iq_j \varphi_\chi(\bm{q}) \varphi_\chi(\bm{k}-\bm{q}) 
+ 8\pi G (\rho+p) q_iq_j v_\chi(\bm{q}) v_\chi(\bm{k}-\bm{q}) 
\bigg]
\, .
\end{equation}
With the scalar solutions given by \eqref{eq:scalarRD-ZSG} and defining
\begin{equation}
\label{eq:RDss-shorthanded}
\frac{q\eta}{\sqrt{3}}
=
\frac{q}{\sqrt{3}k} k\eta
\equiv
\bar{q}_1x
\quad \text{and} \quad
\frac{|{\bm k}-{\bm q}|\eta}{\sqrt{3}}
=
\frac{|{\bm k}-{\bm q}|}{\sqrt{3}k} k\eta
\equiv
\bar{q}_2x 
\, ,
\end{equation}
we can find
\begin{equation}
e_\lambda^{ij}({\bm k}) s_{ij}^{(ss)}(\bm{k}) 
=
\frac{4}{a^2}
\int \frac{d^3q}{(2\pi)^3} \big[ e_\lambda^{ij}({\bm k})q_iq_j \big]
\calR(\bm{q}) \calR(\bm{k}-\bm{q})
\bigg\{
6 \frac{j_1(\bar{q}_1x)}{\bar{q}_1x} \frac{j_1(\bar{q}_2x)}{\bar{q}_2x}
- 2 j_0(\bar{q}_1x) \frac{j_1(\bar{q}_2x)}{\bar{q}_2x}
- 2 \frac{j_1(\bar{q}_1x)}{\bar{q}_1x}  j_0(\bar{q}_2x)
+ j_0(\bar{q}_1x)j_0(\bar{q}_2x)
\bigg\}
\, .
\end{equation}
Thus, the inhomogeneous solution during RD that contains the integral including $G_\text{RD}(\eta,\tilde\eta)$ is
\begin{align}
\label{eq:h2nd-RHS}
h_\lambda(\eta,{\bm k})
& = 
4
\int \frac{d^3q}{(2\pi)^3} \left[ e_\lambda^{ij}({\bm k}) \frac{q_iq_j}{k^2} \right]
\calR(\bm{q}) \calR(\bm{k}-\bm{q})
\int_0^x d\tilde{x}
\tilde{x}^2
\bigg\{
6 \frac{j_1(\bar{q}_1\tilde{x})}{\bar{q}_1\tilde{x}} \frac{j_1(\bar{q}_2\tilde{x})}{\bar{q}_2\tilde{x}}
- 2 j_0(\bar{q}_1\tilde{x}) \frac{j_1(\bar{q}_2\tilde{x})}{\bar{q}_2\tilde{x}}
- 2 \frac{j_1(\bar{q}_1\tilde{x})}{\bar{q}_1\tilde{x}}  j_0(\bar{q}_2\tilde{x})
\nonumber\\
& 
\hspace{22em}
+ j_0(\bar{q}_1\tilde{x})j_0(\bar{q}_2\tilde{x})
\bigg\}
\Big[ j_0(\tilde{x})y_0(x) - j_0(x)y_0(\tilde{x}) \Big]
\, .
\end{align}
The integral over $\tilde{x}$ corresponds to the kernel for the scalar-scalar induced GWs during RD and can be performed analytically, given 
in Appendix~\ref{app:RDss}.

\begin{figure}[ht!]
\begin{center}
\includegraphics[width=0.45\textwidth]{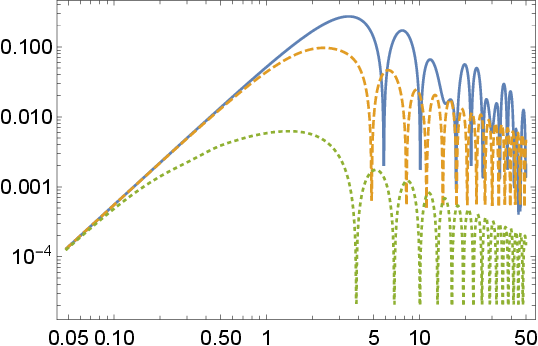}
\hspace{2em}
\includegraphics[width=0.45\textwidth]{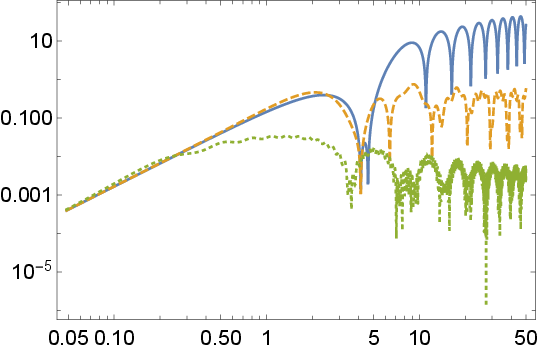}
\end{center}
\caption{The kernels for the scalar-scalar induced GWs during RD in the (left) zero-shear gauge \eqref{eq:h2nd-RHS} and (right) comoving gauge \eqref{eq:h2nd-ssCG} as a function of $k\eta$. Since the kernels are also dependent on $q$, we set (solid lines) $q/k = 0.05$, (dashed lines) $q/k = 2$ and (dotted lines) $q/k = 20$.}
\label{fig:RDss}
\end{figure}

In the comoving gauge, with the same short-handed notations as \eqref{eq:RDss-shorthanded}, straightforward calculations give
\begin{align}
e_\lambda^{ij}({\bm k}) s_{ij}^{(ss)}(\bm{k}) 
& =
\frac{1}{a^2} \int \frac{d^3q}{(2\pi)^3} \big[ e_\lambda^{ij}({\bm k})q_iq_j \big]
\calR(\bm{q}) \calR(\bm{k}-\bm{q})
\bigg\{
j_1(\bar{q}_1x) j_1(\bar{q}_2x) \bigg[ \bar{q}_1\bar{q}_2x^2 + 2 \frac{\bar{q}_2}{\bar{q}_1}
- \frac{2}{\bar{q}_1\bar{q}_2} \big( 1 - 3\bar{q}_1^2 \big) \bigg]
\nonumber\\
&
\hspace{7em}
+ \bigg[ \frac{j_1(\bar{q}_1x)}{\bar{q}_1x} j_0(\bar{q}_2x) 
+ j_0(\bar{q}_1x) \frac{j_1(\bar{q}_2x)}{\bar{q}_2x} \bigg]
x^2 \big( 1 - 3\bar{q}_1^2 + \bar{q}_2^2 \big)
+ j_0(\bar{q}_1x) j_0(\bar{q}_2x) \bigg[ 1 - \frac{x^2}{2} \big( 1 - 3\bar{q}_1^2 + \bar{q}_2^2 \big) \bigg]
\bigg\}
\, ,
\end{align}
so that the solution for $h_\lambda$ is written as
\begin{align}
\label{eq:h2nd-ssCG}
h_\lambda(\eta,{\bm k})
& = 
\int \frac{d^3q}{(2\pi)^3} \left[ e_\lambda^{ij}({\bm k}) \frac{q_iq_j}{k^2} \right]
\calR(\bm{q}) \calR(\bm{k}-\bm{q})
\int_0^x d\tilde{x}
\tilde{x}^2
\bigg\{
j_1(\bar{q}_1\tilde{x}) j_1(\bar{q}_2\tilde{x}) \bigg[ - \bar{q}_1\bar{q}_2\tilde{x}^2 + 2 \frac{\bar{q}_2}{\bar{q}_1}
- \frac{2}{\bar{q}_1\bar{q}_2} \big( 1 - 3\bar{q}_1^2 \big) \bigg]
\nonumber\\
&
\hspace{18em}
+ \bigg[ \frac{j_1(\bar{q}_1\tilde{x})}{\bar{q}_1\tilde{x}} j_0(\bar{q}_2\tilde{x}) 
+ j_0(\bar{q}_1\tilde{x}) \frac{j_1(\bar{q}_2\tilde{x})}{\bar{q}_2\tilde{x}} \bigg]
\tilde{x}^2 \big( 1 - 3\bar{q}_1^2 + \bar{q}_2^2 \big)
\nonumber\\
&
\hspace{18em}
+ j_0(\bar{q}_1\tilde{x}) j_0(\bar{q}_2\tilde{x}) 
\bigg[ 1 - \frac{\tilde{x}^2}{2} \big( 1 - 3\bar{q}_1^2 + \bar{q}_2^2 \big) \bigg]
\bigg\} \Big[ j_0(\tilde{x})y_0(x) - j_0(x)y_0(\tilde{x}) \Big]
\, .
\end{align}
The analytic results for the $\tilde{x}$-integral terms are given in Appendix~\ref{app:RDss}. Comparing \eqref{eq:h2nd-RHS} and \eqref{eq:h2nd-ssCG}, we see that they are clearly different and the gauge dependence of the scalar-induced GWs are persistent during RD.
In Figure~\ref{fig:RDss} we show the kernels in the both gauges.

\subsection{Scalar-tensor induced GWs during RD}

Next we consider the scalar-tensor induced GWs during RD. Again, we work first in the zero-shear gauge for the scalar perturbations. From the linear solutions during RD, the time derivatives of $\varphi_\chi$ and $h_\lambda$ are
\begin{align}
\dot\varphi_\chi({\bm q})
& =
-2H\calR j_2 \bigg( \frac{q\eta}{\sqrt{3}} \bigg) 
\, ,
\\
\ddot\varphi_\chi({\bm q})
& =
2H^2 \calR \bigg[ 5j_2 \bigg( \frac{q\eta}{\sqrt{3}} \bigg) 
- \frac{q\eta}{\sqrt{3}} j_1 \bigg( \frac{q\eta}{\sqrt{3}}\bigg) \bigg]
\, ,
\\
\dot{h}_{\lambda}({\bm q})
&
=
-Hq\eta h_0^{\lambda}({\bm q}) j_1(q\eta)
\, ,
\\
\ddot{h}_\lambda({\bm q})
&
=
\frac{q^2}{a^2} h_0^{\lambda}({\bm q}) j_2(q\eta)
\, .
\end{align}
Then with $q\eta = \bar{q}_1x$ and $|{\bm k}-{\bm q}|\eta = \bar{q}_2x$, we can write
\begin{align}
s_{ij}^{(st)}({\bm k})
& =
\frac{k^2}{a^2}
\int \frac{d^3q}{(2\pi)^3} 
\Bigg\{
\frac{2}{x^2} \bigg[ 2 j_2 \bigg( \frac{\bar{q}_1x}{\sqrt{3}} \bigg) 
- \frac{\bar{q}_1x}{\sqrt{3}} j_1\bigg( \frac{\bar{q}_1x}{\sqrt{3}} \bigg) \bigg] j_0(\bar{q}_2x)
\nonumber\\
& 
\hspace{7em}
+ 2 \bigg( 1 + \frac{|{\bm k}-{\bm q}|^2}{k^2} \bigg) 
\frac{j_1(\bar{q}_1x/\sqrt{3})}{\bar{q}_1x/\sqrt{3}}  j_0(\bar{q}_2x)
\Bigg\} \calR({\bm q}) h^{\lambda'}_{0}({\bm k}-{\bm q})
e_{ij}^{\lambda'}({\bm k}-{\bm q})
\nonumber\\
&
\quad
+
\frac{k^2}{a^2} 
\int \frac{d^3q}{(2\pi)^3} 
\Bigg\{
\frac{2}{x^2} j_0 \bigg( \frac{\bar{q}_1x}{\sqrt{3}} \bigg) \bigg[ 2 j_2(\bar{q}_2x) - \bar{q}_2x j_1(\bar{q}_2x) \bigg] 
+ 2 \bigg( 1 + \frac{q^2}{k^2} \bigg) 
j_0 \bigg( \frac{\bar{q}_1x}{\sqrt{3}} \bigg) \frac{j_1(\bar{q}_2x)}{\bar{q}_2x}  
\Bigg\} \calR({\bm k}-{\bm q}) h^{\lambda'}_{0}({\bm q})
e_{ij}^{\lambda'}({\bm q})
\, .
\end{align}
We note that the first integral can be made identical to the second one upon defining ${\bm k} - {\bm q} \equiv {\bm p}$, and then renaming the dummy integration variable ${\bm p}$ as ${\bm q}$.
Thus, the scalar-tensor source term in the zero-shear gauge is, after multiplying the polarization tensor $e^{ij}_\lambda({\bm k})$,
\begin{align}
e^{ij}_\lambda({\bm k}) s_{ij}^{(st)}({\bm k})
& =
\frac{4k^2}{a^2} 
\int \frac{d^3q}{(2\pi)^3} 
\Big[ e^{ij}_\lambda({\bm k})e_{ij}^{\lambda'}({\bm q}) \Big]
\calR({\bm k}-{\bm q}) h^{\lambda'}_{0}({\bm q})
\nonumber\\
&
\hspace{3em}
\times
\Bigg\{
\frac{1}{x^2} j_0 \bigg( \frac{\bar{q}_1\tilde{x}}{\sqrt{3}} \bigg) 
\bigg[ 2 j_2(\bar{q}_2x) - \bar{q}_2x j_1(\bar{q}_2x) \bigg] 
+ \bigg( 1 + \frac{q^2}{k^2} \bigg) j_0 \bigg( \frac{\bar{q}_1\tilde{x}}{\sqrt{3}} \bigg) 
\frac{j_1(\bar{q}_2x)}{\bar{q}_2x}
\Bigg\} 
\, .
\end{align}
Then the solution is
\begin{align}
\label{eq:hsol-RDscalartensor}
h_\lambda(\eta,{\bm k})
& =
\int \frac{d^3q}{(2\pi)^3} 
\Big[ e^{ij}_\lambda({\bm k})e_{ij}^{\lambda'}({\bm q}) \Big]
\calR({\bm k}-{\bm q}) h^{\lambda'}_{0}({\bm q})
\nonumber\\
&
\quad
\times
4 \int_0^x d\tilde{x} \tilde{x}^2 
\Bigg\{
\frac{1}{\tilde{x}^2} j_0 \bigg( \frac{\bar{q}_1\tilde{x}}{\sqrt{3}} \bigg) 
\bigg[ 2 j_2(\bar{q}_2\tilde{x}) - \bar{q}_2\tilde{x} j_1(\bar{q}_2\tilde{x}) \bigg] 
+ \big( 1 + \bar{q}_1^2 \big) j_0 \bigg( \frac{\bar{q}_1\tilde{x}}{\sqrt{3}} \bigg) 
\frac{j_1(\bar{q}_2\tilde{x})}{\bar{q}_2\tilde{x}} 
\Bigg\} 
\Big[ j_0(\tilde{x})y_0(x) - j_0(x)y_0(\tilde{x}) \Big]
\, .
\end{align}
The $\tilde{x}$-integral can be performed analytically and the 
individual integrations are given in Appendix~\ref{app:RDst}.

In the comoving gauge, similarly upon changing the dummy integration variable we can find the source term as
\begin{align}
e^{ij}_\lambda({\bm k}) s_{ij}^{(st)}({\bm k})
& =
\frac{2k^2}{a^2} 
\int \frac{d^3q}{(2\pi)^3} 
\Big[ e^{ij}_\lambda({\bm k})e_{ij}^{\lambda'}({\bm q}) \Big]
\calR({\bm k}-{\bm q}) h^{\lambda'}_{0}({\bm q})
\bigg[ 2(1-\bar{q}_2^2) j_0\bigg(\frac{\bar{q}_1x}{\sqrt{3}}\bigg) j_0(\bar{q}_2x) 
+ \bigg( -1 + \frac{\bar{q}_1^2}{3} + \bar{q}_2^2 \bigg) \frac{\bar{q}_1x}{\sqrt{3}}
j_1\bigg(\frac{\bar{q}_1x}{\sqrt{3}}\bigg) j_0(\bar{q}_2x) 
\nonumber\\
&
\hspace{20em}
+ 2 \frac{\bar{q}_1^2}{3}\bar{q}_2 x j_0\bigg(\frac{\bar{q}_1x}{\sqrt{3}}\bigg) j_1(\bar{q}_2x) 
+ 2 \frac{\bar{q}_1/\sqrt{3}}{\bar{q}_2} \big( 1 - \bar{q}_1^2 \big) 
j_1\bigg(\frac{\bar{q}_1x}{\sqrt{3}}\bigg) j_1(\bar{q}_2x) 
\bigg]
\, ,
\end{align}
so that the analytic integral solution for $h_\lambda$ is 
\begin{align}
\label{eq:hsol-RDstCG}
h_\lambda(\eta,{\bm k})
& =
\int \frac{d^3q}{(2\pi)^3} 
\Big[ e^{ij}_\lambda({\bm k})e_{ij}^{\lambda'}({\bm q}) \Big]
\calR({\bm k}-{\bm q}) h^{\lambda'}_{0}({\bm q})
\nonumber\\
&
\quad
\times
2 \int_0^x d\tilde{x} \tilde{x}^2 
\bigg[ 2(1-\bar{q}_2^2) j_0\bigg(\frac{\bar{q}_1x}{\sqrt{3}}\bigg) j_0(\bar{q}_2x) 
+ \bigg( -1 + \frac{\bar{q}_1^2}{3} + \bar{q}_2^2 \bigg) \frac{\bar{q}_1x}{\sqrt{3}}
j_1\bigg(\frac{\bar{q}_1x}{\sqrt{3}}\bigg) j_0(\bar{q}_2x) 
\nonumber\\
&
\hspace{7em}
+ 2 \frac{\bar{q}_1^2}{3}\bar{q}_2 x j_0\bigg(\frac{\bar{q}_1x}{\sqrt{3}}\bigg) j_1(\bar{q}_2x) 
+ 2 \frac{\bar{q}_1/\sqrt{3}}{\bar{q}_2} \big( 1 - \bar{q}_1^2 \big) 
j_1\bigg(\frac{\bar{q}_1x}{\sqrt{3}}\bigg) j_1(\bar{q}_2x) 
\bigg]
\Big[ j_0(\tilde{x})y_0(x) - j_0(x)y_0(\tilde{x}) \Big]
\, .
\end{align}
The details of each integration are given in Appendix~\ref{app:RDst}. In Figure~\ref{fig:RDst}, we show both kernels, \eqref{eq:hsol-RDscalartensor} and \eqref{eq:hsol-RDstCG}.

\begin{figure}[ht!]
\begin{center}
\includegraphics[width=0.45\textwidth]{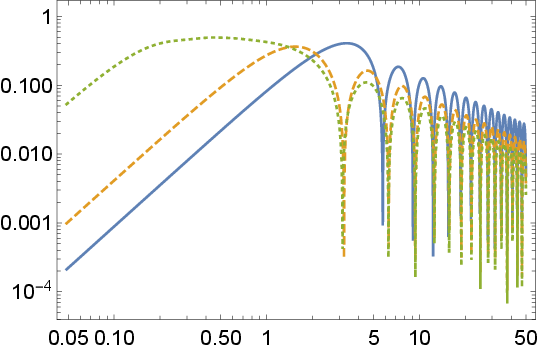}
\hspace{2em}
\includegraphics[width=0.45\textwidth]{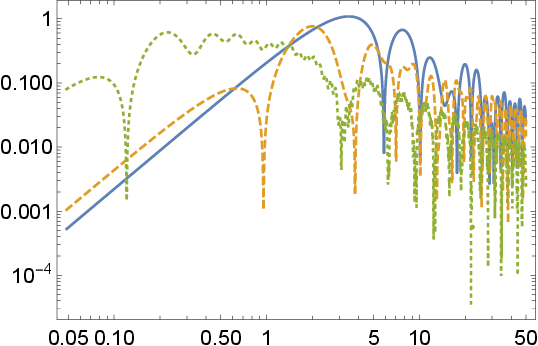}
\end{center}
\caption{The kernels for the scalar-tensor induced GWs during RD in the (left) zero-shear gauge and (right) comoving gauge as a function of $k\eta$. Since the kernel is also dependent on the angle between ${\bm k}$ and ${\bm q}$, for simplicity we set in such a way that for both (solid lines) $q/k = 0.05$ and (dotted lines) $q/k = 20$ they are aligned perpendicular, i.e. $\cos\big(\hat{\bm k}\cdot\hat{\bm q}\big) = 0$, while for (dashed lines) $q/k = 2$ the angle between them is $2\pi/3$, $\cos\big(\hat{\bm k}\cdot\hat{\bm q}\big) = -1/2$.}
\label{fig:RDst}
\end{figure}

\subsection{Tensor-tensor induced GWs during RD}

\begin{figure}[ht!]
\begin{center}
\includegraphics[width=0.45\textwidth]{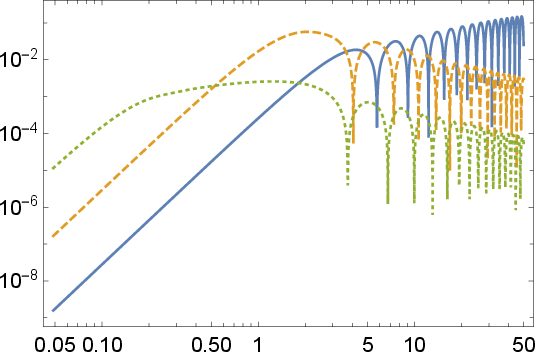}
\hspace{2em}
\includegraphics[width=0.45\textwidth]{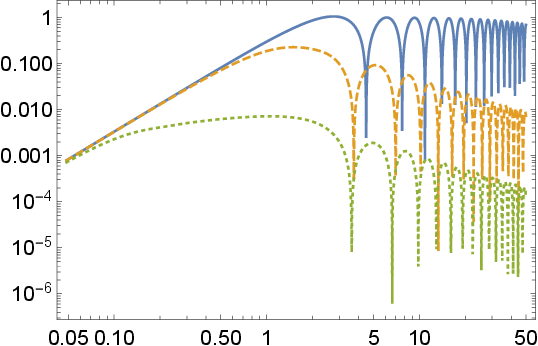}
\end{center}
\caption{The kernels (left) $F_\text{RD}$ and (right) $G_\text{RD}$ as a function of $k\eta$. Since they are also dependent on the angle between ${\bm k}$ and ${\bm q}$, for simplicity we set in such a way that for both (solid lines) $q/k = 0.05$ and (dotted lines) $q/k = 20$ they are aligned perpendicular, i.e. $\cos\big(\hat{\bm k}\cdot\hat{\bm q}\big) = 0$, while for (dashed lines) $q/k = 2$ the angle between them is $2\pi/3$, $\cos\big(\hat{\bm k}\cdot\hat{\bm q}\big) = -1/2$ in both panels.}
\label{fig:RDtt}
\end{figure}

Next we consider the tensor-tensor induced GWs during RD. Working in a similar manner to MD, multiplying the polarization tensor $e^{ij}_\lambda({\bm k})$ gives
\begin{align}
e^{ij}_\lambda({\bm k}) s_{ij}^{(tt)}({\bm k})
& 
=
\frac{1}{a^2}
\int \frac{d^3q_1d^3q_2}{(2\pi)^3} \delta^{(3)}({\bm k}-{\bm q}_{12})
\bigg\{
\underbrace{ h_0^{\lambda_1}({\bm q}_1) h_0^{\lambda_2}({\bm q}_2)
e^{ij}_\lambda({\bm k}) q_1 q_2 e_{i1}^{k\lambda_1} e_{jk2}^{\lambda_2} }_{
\equiv \widetilde{\calC}_\lambda({\bm q}_1,{\bm q}_2)
}
j_1(q_1\eta) j_1(q_2\eta)
\nonumber\\
&
\hspace{10em}
+
h_0^{\lambda_1}({\bm q}_1) h_0^{\lambda_2}({\bm q}_2)
e^{ij}_\lambda({\bm k})
\bigg[
e^{kl}_{\lambda_11} \big(
-q_{2j}q_{2k} e^{\lambda_2}_{il2} - q_{2i}q_{2k} e_{jl2}^{\lambda_2} 
+ q_{2k}q_{2l} e^{\lambda_2}_{ij2} 
\big)
\nonumber\\
&
\hspace{11em}
\underbrace{ 
\hspace{8em}
+ \frac{1}{2} q_{1i}q_{1j} e_{\lambda_1}^{kl1} e_{kl2}^{\lambda_2} 
+ ({\bm q}_1\cdot{\bm q}_2) e_{i1}^{k\lambda_1} e_{jk2}^{\lambda_2}
- q_1^lq_{2k} e_{i1}^{k\lambda_1} e_{jl2}^{\lambda_2} 
\bigg] }_{
\equiv \widetilde{\calD}_\lambda({\bm q}_1,{\bm q}_2)
}
j_0(q_1\eta) j_0(q_2\eta)
\bigg\}
\nonumber\\
& 
\quad
+ ({\bm q}_1 \leftrightarrow {\bm q}_2)
\, .
\end{align}
Thus, with $x=k\eta$, $\bar{q}_1 \equiv q_1/k$ and $\bar{q}_2 \equiv q_2/k$,
\begin{align}
h_\lambda(\eta,{\bm k})
& =
\int \frac{d^3q_1d^3q_2}{(2\pi)^3} \delta^{(3)}({\bm k}-{\bm q}_{12}) \frac{1}{k^2} \bigg\{
\Big[ \widetilde{\calC}_\lambda({\bm q}_1,{\bm q}_2) + ({\bm q}_1 \leftrightarrow {\bm q}_2) \Big]
\int_0^x d\tilde{x} \tilde{x}^2
j_1(\bar{q}_1\tilde{x}) j_1(\bar{q}_2\tilde{x})
\Big[ j_0(\tilde{x})y_0(x) - j_0(x)y_0(\tilde{x}) \Big]
\nonumber\\
& 
\hspace{13.5em}
+
\Big[ \widetilde{\calD}_\lambda({\bm q}_1,{\bm q}_2) + ({\bm q}_1 \leftrightarrow {\bm q}_2) \Big]
\int_0^x d\tilde{x} \tilde{x}^2
j_0(\bar{q}_1\tilde{x}) j_0(\bar{q}_2\tilde{x})
\Big[ j_0(\tilde{x})y_0(x) - j_0(x)y_0(\tilde{x}) \Big] 
\bigg\}
\, .
\end{align}
The $\tilde{x}$-integrals can be performed analytically and finally we can write
\begin{align}
\label{eq:tt-source-sol2}
h_\lambda({\bm k},\eta) 
& =
\int \frac{d^3q}{(2\pi)^3} h_0^{\lambda_1}({\bm q}) h_0^{\lambda_2}({\bm k}-{\bm q})
\Bigg\{
e_\lambda^{ij}({\bm k}) e_i^{k\lambda_1}({\bm q}) e_{jk}^{\lambda_2}({\bm k}-{\bm q}) 
{F}_\text{RD}({\bm k},{\bm q},\eta)
\nonumber\\
& 
\hspace{1.5em}
+
\frac{1}{k^2} 
\bigg(
e_\lambda^{ij}({\bm k})
e^{kl}_{\lambda_1}({\bm q}) \Big(
q_jk_k e_{il}^{\lambda_2}({\bm k}-{\bm q}) 
+ q_ik_k e_{jl}^{\lambda_2}({\bm k}-{\bm q})
+ k_kk_l e_{ij}^{\lambda_2}({\bm k}-{\bm q})
\Big)
\nonumber\\
&
\hspace{2em}
+
e_\lambda^{ij}({\bm k}) 
\bigg[
\frac{1}{2} q_i q_j e^{kl}_{\lambda_1}({\bm q}) e_{kl}^{\lambda_2}({\bm k}-{\bm q})
+ {\bm q}\cdot({\bm k}-{\bm q}) e_i^{k\lambda_1}({\bm q}) e_{jk}^{\lambda_2}({\bm k}-{\bm q})
- q^lk_k e_i^{k\lambda_1}({\bm q}) e_{jl}^{\lambda_2}({\bm k}-{\bm q})
\bigg]
\bigg) 
{G}_\text{RD}({\bm k},{\bm q},\eta)
\Bigg\}
\, ,
\end{align}
where the kernels $F_\text{RD}$ and $G_\text{RD}$ are given respectively by \eqref{eq:RDttF} and \eqref{eq:RDttG}. In Figure~\ref{fig:RDtt} we show $F_\text{RD}$ and $G_\text{RD}$.

\section{Conclusions}
\label{sec:conc}

In this article, we have presented the equation of motion for the tensor perturbations up to second order in perturbations, including all possible quadratic combinations of different types of cosmological perturbations. These terms serve as sources to generate second-order GWs. Given that the universe is filled with a perfect fluid matter with vanishing anisotropic stress, only linear scalar and tensor perturbations contribute to the source terms. And we have found the analytic integral solutions for the second-order GWs during both MD and RD induced by the scalar-scalar, scalar-tensor and tensor-tensor sources. The transition between MD and RD can be considered by separating the time integral of \eqref{eq:generalsol}. That is, for the transition from the epoch $A$ to the epoch $B$ at $\eta = \eta_\star$, we may write~\citep{2018PhRvD..97l3532K}
\begin{equation}
h_\lambda(\eta,{\bm k})
=
\frac{1}{a} \Bigg\{
\int_0^{\eta_\star} d\tilde\eta \, \frac{a^3(\eta)}{a^3(\eta_\star)} 
\Big[ a^3(\tilde\eta)  e^{ij}_\lambda(\bm{k}) s^A_{ij}(\bm{k}) \Big]
G_{A\to B}(\eta,\tilde\eta)
+
\int_{\eta_\star}^\eta d\tilde\eta \, \Big[ a^3(\tilde\eta)  e^{ij}_\lambda(\bm{k}) s^{A\to B}_{ij}(\bm{k}) \Big]
G_B(\eta,\tilde\eta)
\Bigg\}
\, ,
\end{equation}
where the first term represents the change of the propagation of the GWs produced during $A$ through $B$, and the second term denotes the modification of the source. These terms can be found by matching, for the first term, the solutions and, for the second term, the kernels. However given the complicated source terms, it is a formidable task to compute the effects of the transition analytically, so we do not proceed any further but are satisfied with the above schematic form, which can in principle be further manipulated.

Since the primary tensor perturbations are persistent irrespective of the sources, it is interesting to discuss if the tensor-induced GWs could ever be observationally significant. To address this question quantitatively, it is necessary to study further the relation between the tensor-induced GWs and the primary GWs. For example, if the primary GWs exhibit a sharp peak around at a certain scale $k$, the corresponding tensor-induced GWs are likely to be more prominent than the primary ones over a broader range of the wavenumber other than the peak. In this regards, the tensor-induced GWs may serve as a useful tool to probe the shape of the primary GWs and the relevant underlying physics. It is also interesting to consider how to test the origin of the induced GWs. We may consider, for example, the scalar-scalar-tensor bispectrum, $\langle\calR\calR h\rangle$. If the tensor mode is scalar-induced, i.e. $h \sim \calR\calR$, the bispectrum is directly related to the local-type scalar bispectrum. If, on the other hand, the tensor mode is induced by the primary GWs, $\langle\calR\calR h\rangle$ is proportional to the product of the scalar and tensor power spectra. Higher-order correlation functions may thus be able to serve as a possible tool to test the origin of the induced tensor perturbations.

As the kernels that involve rapid oscillations are analytically specified, the solutions given in this article should be useful for analytic and/or numerical studies of the second-order induced GWs.

\subsection*{Acknowledgements}

I thank Jai-chan Hwang, Donghui Jeong, Kazunori Kohri, Sachiko Kuroyanagi, Takahiro Terada and Jaiyul Yoo for helpful comments and discussions. I also thank the anonymous referee for her/his constructive feedback that improved the content of this article. 
This work is supported in part by the Basic Science Research Program (2016R1D1A1B03930408) and Mid-career Research Program (2019R1A2C2085023) through the National Research Foundation of Korea Research Grants. 
I also acknowledge the Korea-Japan Basic Scientific Cooperation Program supported by the National Research Foundation of Korea and the Japan Society for the Promotion of Science (2018K2A9A2A08000127 and 2020K2A9A2A08000097) and the Ewha Womans University Research Grant of 2020 (1-2020-1630-001-1). 
I am grateful to the Asia Pacific Center for Theoretical Physics for Focus Research Program ``The origin and evolution of the Universe'' where parts of this work were presented and discussed.

\newpage

\appendix

\renewcommand{\theequation}{\Alph{section}.\arabic{equation}}

\section{Traceless evolution equation for the spatial metric}
\label{app:traceless-eq}
\setcounter{equation}{0}

The traceless evolution equation for the spatial metric is written compactly using the Arnowitt-Deser-Misner formulation~\citep{2008GReGr..40.1997A}. With the metric
\begin{equation}
ds^2 = -N^2(dx^0)^2 + \gamma_{ij} (N^idx^0+dx^i) (N^jdx^0+dx^j) \, ,
\end{equation}
where $N$, $N^i$ and $\gamma_{ij}$ denote respectively the lapse function, shift vector and spatial metric, the dynamics of the space-time is described by the spatial metric $\gamma_{ij}$ through the curvature variables of the spatial hypersurfaces. Along with the matter contents residing in the space-time, the geometric equations for the curvature variables constitute a complete set of the equations of motion.
The extrinsic curvature $K_{ij}$ is introduced as
\begin{equation}
K_{ij} \equiv \frac{1}{2N} \big( N_{i|j} + N_{j|i} - \gamma_{ij,0} \big) \, ,
\end{equation}
where a vertical bar denotes a covariant derivative with respect to $\gamma_{ij}$. The evolution equation for the traceless part $\overline{K}_{ij} = K_{ij} - \gamma_{ij}K/3$ with $K \equiv K^i{}_i$ is
\begin{equation}
\frac{\overline{K}^i{}_{j,0}}{N} - \frac{\overline{K}^i{}_{j|k}N^k}{N} 
+ \frac{\overline{K}^k{}_jN^i{}_{|k}}{N} - \frac{\overline{K}^i{}_kN^k{}_{|j}}{N} 
= 
K\overline{K}^i{}_j - \frac{1}{N} \left( N^{|i}{}_{|j} 
- \frac{\delta^i{}_j}{3}N^{|k}{}_{|k} \right) + \overline{R}^i{}_j
- 8\pi G \overline{T}^i{}_j \, ,
\end{equation}
where $\overline{R}_{ij}$ is the traceless part of the intrinsic curvature tensor $R_{ij}$ constructed from $\gamma_{ij}$. This is fully non-linear, geometric equation. To incorporate cosmological perturbations, we expand it perturbatively up to desired accuracy~\citep{2004PhRvD..69j4011N,2007PhRvD..76j3527H}, or write the exact non-linear equation for cosmological perturbations~\citep{2017JCAP...10..027G}. Using \eqref{eq:metric} and expanding up to second order in perturbations, this equation is written as 

\begin{align}
\label{eq:traceless-eq}
&
\ddot{h}^i{}_j + 3H\dot{h}^i{}_j - \frac{\Delta}{a^2}h^i{}_j 
+ \frac{1}{a^2} \left\{ \left[ 
a^2 \frac{d}{dt} \left( \frac{\chi^{(i}{}_{,j)}}{a^2} \right) + 3H \chi^{(i}{}_{,j)} 
\right] 
- \frac{\delta^i{}_j}{3} \left[ 
a^2 \frac{d}{dt} \left( \frac{\chi^k{}_{,k}}{a^2} \right) + 3H \chi^k{}_{,k} 
\right] \right\}
\nonumber\\
&
- \frac{1}{a^2} \bigg( \partial^i\partial_j - \frac{\delta^i{}_j}{3}\Delta \bigg)
(\alpha + \varphi)
- 8\pi G \bigg( \Pi^i{}_j - \frac{\delta^i{}_j}{3}\Pi^k{}_k \bigg)
\nonumber\\
= & 
\frac{d}{dt} \left[ 
\left( \dot{h}^i{}_j + \frac{\chi^{(i}{}_{,j)}}{a^2} \right) \alpha
+ 2 \big( \varphi\delta^{ik} + h^{ik} \big) \bigg( \dot{h}_{jk} + \frac{\chi_{(k,j)}}{a^2} \bigg) + 2h^{ik}\dot\varphi\delta_{jk}
+ \frac{\chi^k}{a^2} \Big( \varphi^{,i}\delta_{jk} + \varphi_{,j}\delta^i{}_k - \varphi_{,k}\delta^i{}_j
+ h^i{}_{k,j} + h_{jk}{}^{,i} - h^i{}_{j,k} \Big) 
\right]
\nonumber\\
&
+ 3H \left[
\left( \dot{h}^i_j + \frac{\chi^{(i}{}_{,j)}}{a^2} \right) \alpha
+ 2 \big( \varphi\delta^{ik} + h^{ik} \big) \bigg( \dot{h}_{jk} + \frac{\chi_{(k,j)}}{a^2} \bigg) + 2h^{ik}\dot\varphi\delta_{jk}
+ \frac{\chi^k}{a^2} \Big( \varphi^{,i}\delta_{jk} + \varphi_{,j}\delta^i{}_k - \varphi_{,k}\delta^i{}_j
+ h^i{}_{k,j} + h_{jk}{}^{,i} - h^i{}_{j,k} \Big) 
\right]
\nonumber\\
&
+ \alpha \frac{d}{dt} \left( \dot{h}^i{}_j + \frac{\chi^{(i}{}_{,j)}}{a^2} \right)
- \frac{\chi^k}{a^2} \bigg( \dot{h}^i{}_{j,k} + \frac{1}{a^2}\chi^{(i}{}_{,j)k} \bigg)
+ \kappa \bigg( \dot{h}^i{}_j + \frac{1}{a^2} \chi^{(i}{}_{,j)} \bigg)
\nonumber\\
&
- \frac{\delta^i{}_j}{3} \Bigg\{
\frac{d}{dt} \left[ 
\frac{\chi^k{}_{,k}}{a^2}\alpha 
+ 2 \big( \varphi\delta^{kl} + h^{kl} \big) \frac{\chi_{(k,l)}}{a^2} 
+ 2h^{kl}\dot{h}_{kl} - \frac{1}{a^2}\chi^k\varphi_{,k} 
\right]
+ 3H \left[ 
\frac{\chi^k{}_{,k}}{a^2}\alpha 
+ 2 \big( \varphi\delta^{kl} + h^{kl} \big) \frac{\chi_{(k,l)}}{a^2} 
+ 2h^{kl}\dot{h}_{kl} - \frac{1}{a^2}\chi^k\varphi_{,k} 
\right]
\nonumber\\
& 
\hspace{3em}
+ \alpha \frac{d}{dt} \left( \frac{\chi^k{}_{,k}}{a^2} \right)
- \frac{\chi^k}{a^2} \frac{1}{a^2}\chi^l{}_{,lk}
+ \kappa \frac{\chi^k{}_{,k}}{a^2}
\Bigg\}
+ \frac{1}{a^2}\chi^i{}_{,k} \bigg( \dot{h}^k{}_j + \frac{1}{a^2}\chi^{(k}{}_{,j} \bigg)
-\frac{1}{a^2} \chi^k{}_{,j} \bigg( \dot{h}^i{}_k + \frac{1}{a^2} \chi^{(i}{}_{,k} \bigg)
\nonumber\\
&
+ \frac{1}{a^2} \left\{ 
- \alpha\alpha^{,i}{}_{,j}
+ \frac{1}{2} \bigg( -\alpha^2 + \frac{1}{a^2}\chi^l\chi_l \bigg)^{,i}{}_{,j}
- 2 \big( \varphi\delta^{ik} + h^{ik} \big) \alpha_{,jk}  
- \Big( 
\varphi_{,j}\delta^i{}_k  + h^i{}_{k,j} + \varphi^{,i}\delta_{jk} + h_{jk}{}^{,i}
- \varphi_{,k}\delta^i{}_j - h^i{}_{j,k} \Big) \alpha^{,k} 
\right\}
\nonumber\\
& -
\frac{\delta^i{}_j}{3} \left\{
\frac{1}{a^2} \bigg[
- \alpha\Delta\alpha
+ \frac{1}{2}\Delta \bigg( -\alpha^2 + \frac{1}{a^2} \chi^k\chi_k \bigg)
- 2 \big( \varphi\delta^{kl} + h^{kl} \big) \alpha_{,kl}
+ \varphi^{,k}\alpha_{,k}
\bigg] \right\}
\nonumber\\
&
+ \frac{1}{a^2} \bigg[ 
- 3\varphi^{,i}\varphi_{,i} - 4\varphi\varphi^{,i}{}_{,j}  
- 4\varphi\Delta{h}^i{}_j - 2h^i{}_j\Delta\varphi + \varphi_{,k}h^{ik}{}_{,j} + \varphi^{,k}h_{jk}{}^{,i} - 3\varphi^{,k}h^i{}_{j,k}
+ 2\varphi^{,ik}h_{jk} 
\nonumber\\
& \qquad
- 2h^{ik}\Delta{h}_{kj} - 2h^{kl}h_{kl,j}{}^{,i}
- h^{kl,i}h_{kl,j} - 2h^{ik,l}h_{jk,l}  
+ 2h^{ik,l}h_{jl,k} - 2h^{kl} \Big( h^i{}_{l,jk} + h_{jl,k}{}^{,i} - h^i{}_{j,kl} \Big) 
\nonumber\\
& \qquad
- \frac{\delta^i{}_j}{3} \Big(
-3\varphi^{,k}\varphi_{,k} - 4\varphi\Delta\varphi + 2\varphi^{,kl}h_{,kl}
- 4h^{kl}\Delta{h}_{kl} - 3h^{kl,m}h_{kl,m} + 2h^{kl,m}h_{km,l} \Big) 
\bigg]
\nonumber\\
&
- 16\pi G \bigg[
\Big( \varphi\delta^{ik} + h^{ik} \Big) \Pi_{jk}
- \frac{\delta^i{}_j}{3} \big( \varphi\delta^{kl} + h^{kl} \big)\Pi_{kl}
\bigg]
+ 8\pi G (\rho+p) \bigg( v^iv_j - \frac{\delta^i{}_j}{3}v^kv_k \bigg)
\, .
\end{align}

\section{Integrals of Bessel functions}
\label{app:integrals}
\setcounter{equation}{0}

Here, $a$, $b$ and $c$ denote arbitrary positive constants.

\subsection{Integrals for scalar-scalar and scalar-tensor induced GWs during MD}
\label{app:MDss-st}

For $s_{ij}^{(ss)}$ during MD, we have for the zero-shear gauge the following integrals for the spherical Bessel functions:
\begin{align}
\int_0^x d\tilde{x} \tilde{x}^3 j_1(\tilde{x}) 
& = 
3x^2j_1(x) - x^3j_0(x)
\, ,
\\
\int_0^x d\tilde{x} \tilde{x}^3 y_1(\tilde{x}) 
& = 
3 + 3x^2y_1(x) - x^3y_0(x) 
\, ,
\end{align}
along with the following identities of the spherical Bessel functions:
\begin{align}
\label{eq:sphB-identity}
\bigg( \frac{1}{x} \frac{d}{dx} \bigg)^m \Big[ x^{n+1}f_n(x) \Big] 
& = 
x^{n-m+1}f_{n-m}(x)
\, ,
\\
\label{eq:recurrence}
f_{n-1}+f_{n+1} 
& = 
\frac{2n+1}{x}f_n
\, .
\end{align}
Also, we have the following integrals for the comoving gauge:
\begin{align}
\int_0^x d\tilde{x} \tilde{x}^5 j_1(\tilde{x}) 
& = 
5 x^2 ( x^2 - 6 ) j_1(x) - x^3 ( x^2 - 10 ) j_0(x)
\, ,
\\
\int_0^x d\tilde{x} \tilde{x}^5 y_1(\tilde{x}) 
& = 
- 30 + 5x^2 ( x^2 - 6 ) y_1(x) - x^3 ( x^2 - 10 ) y_0(x)
\, .
\end{align}

For $s_{ij}^{(st)}$ during MD, for the zero-shear gauge, 
\begin{align}
\int_0^x d\tilde{x} \tilde{x}^2 j_1(a\tilde{x}) j_1(\tilde{x})
& =
\frac{1}{a^2-1} \Big[ x^2 j_1(ax) j_0(x) - ax^2j_0(ax)  j_1(x) \Big]
\, ,
\\
\int_0^x d\tilde{x} \tilde{x}^2 j_1(a\tilde{x}) y_1(\tilde{x})
& =
\frac{1}{a^2-1} \Big[ - a + x^2 j_1(ax) y_0(x) - ax^2 j_0(ax) y_1(x) \Big]
\, .
\end{align}
These give a rather simple result:
\begin{align}
\label{eq:st-integral}
\int_0^x d\tilde{x} \tilde{x}^2 j_1(a\tilde{x}) \Big[ j_1(\tilde{x}) y_1(x) - j_1(x) y_1(\tilde{x}) \Big]
& = 
\frac{1}{a^2-1} 
\Big[ aj_1(x) - j_1(ax) \Big]
\, .
\end{align}
For the comoving gauge, we have the following integrals:
\begin{align}
\int_0^x d\tilde{x} \tilde{x}^3 j_0(a\tilde{x}) j_1(\tilde{x})
& =
\frac{x^2}{(a^2-1)^2} \Big\{ j_0(x) \big[ (a^2-1)xj_0(ax) - 2aj_1(ax) \big]
+ j_1(x) \big[ -(a^2-3)j_0(ax) + (a^2-1)axj_1(ax) \big] \Big\}
\, ,
\\
\int_0^x d\tilde{x} \tilde{x}^3 j_0(a\tilde{x}) y_1(\tilde{x})
& =
\frac{1}{(a^2-1)^2} \Big\{ 3 - a^2 + x^2j_0(ax) \big[ (a^2-1)xy_0(x) - (a^2-3)y_1(x) \big]
\nonumber\\
&
\hspace{6em}
+ ax^2j_1(ax) \big[ -2y_0(x) + (a^2-1)xy_1(x) \big] \Big\}
\, .
\end{align}
Again, we find a rather simple result from these integrals:
\begin{align}
\int_0^x d\tilde{x} \tilde{x}^3 j_0(a\tilde{x}) \Big[ j_1(\tilde{x}) y_1(x) - j_1(x) y_1(\tilde{x}) \Big]
& =
\frac{1}{(a^2-1)^2} \Big[ (a^2-3) j_1(x) + 2aj_1(ax) - (a^2-1)xj_0(ax) \Big]
\, .
\end{align}

\subsection{Integrals for tensor-tensor induced GWs during MD}
\label{app:integrals-MDtt}

They can be arranged as, for the first two integrals,
\begin{align}
& \int_0^x d\tilde{x} \tilde{x} j_2(a\tilde{x}) j_2(b\tilde{x}) j_1(c\tilde{x})
\nonumber\\
& =
\frac{(a^2+b^2-c^2)(a^2-b^2-c^2)x^2}{16ab^2c} j_1(ax)j_0(bx)j_0(cx)
- \frac{(a^2+b^2-c^2)(a^2-b^2+c^2)x^2}{16a^2bc} j_0(ax)j_1(bx)j_0(cx)
\nonumber\\
&
\quad
- \frac{(a^2+b^2-c^2)^2x^2}{16a^2b^2} j_0(ax)j_0(bx)j_1(cx)
+ \frac{x}{2a} j_1(ax)j_0(bx)j_1(cx) + \frac{x}{2b} j_0(ax)j_1(bx)j_1(cx)
- \frac{cx}{2ab} j_1(ax)j_1(bx)j_0(cx)
\nonumber\\
&
\quad
- \frac{(a^2+b^2-c^2)x^2+12}{8ab} j_1(ax)j_1(bx)j_1(cx)
\nonumber\\
&
\quad
+ \frac{(a-b-c)(a+b-c)(a-b+c)(a+b+c)(a^2+b^2-c^2)}{64a^3b^3c^2}
\nonumber\\
&
\qquad
\times
\Big\{ {\rm Si}[(a-b-c)x] - {\rm Si}[(a+b-c)x]
- {\rm Si}[(a-b+c)x] + {\rm Si}[(a+b+c)x] \Big\}
\, ,
\\
& \int_0^x d\tilde{x} \tilde{x} j_2(a\tilde{x}) j_2(b\tilde{x}) y_1(c\tilde{x})
\nonumber\\
& =
\frac{(a^2+b^2-c^2)(a^2-b^2-c^2)x^2}{16ab^2c} j_1(ax)j_0(bx)y_0(cx)
- \frac{(a^2+b^2-c^2)(a^2-b^2+c^2)x^2}{16a^2bc} j_0(ax)j_1(bx)y_0(cx)
\nonumber\\
&
\quad
- \frac{(a^2+b^2-c^2)^2x^2}{16a^2b^2} j_0(ax)j_0(bx)y_1(cx)
+ \frac{x}{2a} j_1(ax)j_0(bx)y_1(cx) + \frac{x}{2b} j_0(ax)j_1(bx)y_1(cx)
- \frac{cx}{2ab} j_1(ax)j_1(bx)y_0(cx)
\nonumber\\
&
\quad
- \frac{(a^2+b^2-c^2)x^2+12}{8ab} j_1(ax)j_1(bx)y_1(cx)
\nonumber\\
&
\quad
- \frac{(a-b-c)(a+b-c)(a-b+c)(a+b+c)(a^2+b^2-c^2)}{64a^3b^3c^2}
\nonumber\\
&
\qquad
\times
\Big\{ {\rm Cin}[(a-b-c)x] - {\rm Cin}[(a+b-c)x]
+ {\rm Cin}[(a-b+c)x] - {\rm Cin}[(a+b+c)x] \Big\}
\nonumber\\
&
\quad
- \frac{3a^4+3(b^2-c^2)^2-2a^2(b^2+3c^2)}{48a^2b^2c^2}
\, .
\end{align}
Note that with the definition
\begin{equation}
{\rm Cin}(x) \equiv \int_0^x \frac{1-\cos{t}}{t} dt = \gamma + \log{x} - {\rm Ci}(x) 
\, ,
\end{equation}
where $\gamma \approx 0.577216$ is the Euler-Mascheroni constant, while both ${\rm Ci}(x)$ and $\log{x}$ are diverging as $x\to0$, $\lim_{x\to0}{\rm Cin}(x) = 0$. Then we find  
\begin{align}
\label{eq:j2j2-integral}
&
\frac{1}{x} \int_0^x d\tilde{x} \tilde{x} j_2(a\tilde{x}) j_2(b\tilde{x})
\Big[ j_1(\tilde{x})y_1(x) - j_1(x)y_1(\tilde{x}) \Big]
\equiv 
F_\text{MD}(a,b,x)
\nonumber\\
& =
\frac{(a^2+b^2-1)(-a^2+b^2+1)}{16b^2} \frac{j_1(ax)}{ax} j_0(bx)
+ \frac{(a^2+b^2-1)(a^2-b^2+1)}{16a^2} j_0(ax) \frac{j_1(bx)}{bx}
+ \frac{1}{2} \frac{j_1(ax)}{ax} \frac{j_1(bx)}{bx}
\nonumber\\
& 
\quad
+ \frac{3a^4+3b^4+3 - 6a^2-6b^2 - 2a^2b^2}{48a^2b^2} \frac{j_1(x)}{x}
\nonumber\\
&
\quad
+ \frac{(a-b-1)(a+b-1)(a-b+1)(a+b+1)(a^2+b^2-1)}{64a^3b^3} 
\nonumber\\
&
\qquad
\times
\bigg\{ \bigg( {\rm Si}[(a-b-1)x] \frac{y_1(x)}{x} + {\rm Cin}[(a-b-1)x] \frac{j_1(x)}{x} \bigg) 
- \bigg( {\rm Si}[(a+b-1)x] \frac{y_1(x)}{x} + {\rm Cin}[(a+b-1)x] \frac{j_1(x)}{x} \bigg) 
\nonumber\\
& 
\hspace{4em}
- \bigg( {\rm Si}[(a-b+1)x] \frac{y_1(x)}{x} - {\rm Cin}[(a-b+1)x] \frac{j_1(x)}{x} \bigg) 
+ \bigg( {\rm Si}[(a+b+1)x] \frac{y_1(x)}{x} - {\rm Cin}[(a+b+1)x] \frac{j_1(x)}{x} \bigg) \bigg\}
\, .
\end{align}

Likewise, for the next two integrals,
\begin{align}
& \int_0^x d\tilde{x} \tilde{x} j_1(a\tilde{x}) j_1(b\tilde{x}) j_1(c\tilde{x})
\nonumber\\
& =
-\frac{x^2}{4} j_1(ax)j_1(bx)j_1(cx)
\nonumber\\
&
\quad
- \frac{(a^2+b^2-c^2)x^2}{8ab} j_0(ax)j_0(bx)j_1(cx)
- \frac{(-a^2+b^2+c^2)x^2}{8ab} j_1(ax)j_0(bx)j_0(cx)
- \frac{(a^2-b^2+c^2)x^2}{8ab} j_0(ax)j_1(bx)j_0(cx)
\nonumber\\
&
\quad
+ \frac{(a-b-c)(a+b-c)(a-b+c)(a+b+c)}{32a^2b^2c^2}
\Big\{ {\rm Si}[(a-b-c)x] - {\rm Si}[(a+b-c)x]
- {\rm Si}[(a-b+c)x] + {\rm Si}[(a+b+c)x] \Big\}
\, ,
\\
& \int_0^x d\tilde{x} \tilde{x} j_1(a\tilde{x}) j_1(b\tilde{x}) y_1(c\tilde{x})
\nonumber\\
& =
-\frac{x^2}{4} j_1(ax)j_1(bx)y_1(cx)
\nonumber\\
&
\quad
- \frac{(a^2+b^2-c^2)x^2}{8ab} j_0(ax)j_0(bx)y_1(cx)
- \frac{(-a^2+b^2+c^2)x^2}{8ab} j_1(ax)j_0(bx)y_0(cx)
- \frac{(a^2-b^2+c^2)x^2}{8ab} j_0(ax)j_1(bx)y_0(cx)
\nonumber\\
&
\quad
- \frac{(a-b-c)(a+b-c)(a-b+c)(a+b+c)}{32a^2b^2c^2}
\Big\{ {\rm Cin}[(a-b-c)x] - {\rm Cin}[(a+b-c)x]
+ {\rm Cin}[(a-b+c)x] - {\rm Cin}[(a+b+c)x] \Big\}
\nonumber\\
&
\quad
- \frac{a^2+b^2-c^2}{8abc^2}
\, .
\end{align}
Thus,
\begin{align}
\label{eq:j1j1-integral}
&
\frac{1}{abx} \int_0^x d\tilde{x} \tilde{x} j_1(a\tilde{x}) j_1(b\tilde{x})
\Big[ j_1(\tilde{x})y_1(x) - j_1(x)y_1(\tilde{x}) \Big]
\equiv
G_\text{MD}(a,b,x)
\nonumber\\
& =
\frac{(-a^2+b^2+1)}{8b^2} \frac{j_1(ax)}{ax} j_0(bx) 
+ \frac{(a^2-b^2+1)}{8a^2} j_0(ax) \frac{j_1(bx)}{bx}
+ \frac{a^2+b^2-1}{8a^2b^2} \frac{j_1(x)}{x}
\nonumber\\
&
\quad
+ \frac{(a-b-1)(a+b-1)(a-b+1)(a+b+1)}{32a^3b^3}
\nonumber\\
&
\qquad
\times
\bigg\{ \bigg( {\rm Si}[(a-b-1)x] \frac{y_1(x)}{x} + {\rm Cin}[(a-b-1)x] \frac{j_1(x)}{x} \bigg) 
- \bigg( {\rm Si}[(a+b-1)x] \frac{y_1(x)}{x} + {\rm Cin}[(a+b-1)x] \frac{j_1(x)}{x} \bigg) 
\nonumber\\
& 
\hspace{3em}
- \bigg( {\rm Si}[(a-b+1)x] \frac{y_1(x)}{x} - {\rm Cin}[(a-b+1)x] \frac{j_1(x)}{x} \bigg) 
+ \bigg( {\rm Si}[(a+b+1)x] \frac{y_1(x)}{x} - {\rm Cin}[(a+b+1)x] \frac{j_1(x)}{x} \bigg) \bigg\}
\, .
\end{align}
Notice that comparing with \eqref{eq:j2j2-integral}, $F_\text{MD}(a,b,x)$ and $G_\text{MD}(a,b,x)$ are related by
\begin{equation}
\frac{a^2+b^2-1}{2}G_\text{MD}(a,b,x)
=
F_\text{MD}(a,b,x) - \frac{1}{2} \frac{j_1(ax)}{ax} \frac{j_1(bx)}{bx}
+ \frac{1}{6} \frac{j_1(x)}{x} 
\, .
\end{equation}

\subsection{Integrals for scalar-scalar induced GWs during RD}
\label{app:RDss}

We find
\begin{align}
\int_0^x d\tilde{x} \tilde{x}^2 j_0(a\tilde{x})j_0(b\tilde{x})j_0(c\tilde{x})
& =
-\frac{1}{4abc} \Big\{
{\rm Si}\big[ (a-b-c)x\big] - {\rm Si}\big[ (a+b-c)x\big] - {\rm Si}\big[ (a-b+c)x\big] + {\rm Si}\big[ (a+b+c)x\big]
\Big\}
\, ,
\\
\int_0^x d\tilde{x} \tilde{x}^2 j_0(a\tilde{x})j_0(b\tilde{x})y_0(c\tilde{x})
& =
\frac{1}{4abc} \Big\{
{\rm Cin}\big[(a-b-c)x\big] - {\rm Cin}\big[(a+b-c)x\big] + {\rm Cin}\big[(a-b+c)x\big] - {\rm Cin}\big[(a+b+c)x\big]
\Big\}
\, ,
\end{align}
\begin{align}
&
\int_0^x d\tilde{x}
\tilde{x}
j_1(a\tilde{x})j_0(b\tilde{x}) j_0(c\tilde{x})
\nonumber\\
&
=
-\frac{x}{a} j_0(ax) j_0(bx) j_0(cx)
- \frac{1}{2}x^2 j_1(ax) j_0(bx) j_0(cx)
+ \frac{b}{2a}x^2 j_0(ax) j_1(bx) j_0(cx)
+ \frac{c}{2a}x^2 j_0(ax) j_0(bx) j_1(cx)
\nonumber\\
&
\quad
+ \frac{1}{8a^2bc} \Big\{
- \big( a^2-(b+c)^2 \big) {\rm Si}[(a-b-c)x] + \big( a^2-(b-c)^2 \big) {\rm Si}[(a+b-c)x]
\nonumber\\
&
\quad
\hspace{4em}
+ \big( a^2-(b-c)^2 \big) {\rm Si}[(a-b+c)x] - \big( a^2-(b+c)^2 \big) {\rm Si}[(a+b+c)x]
\Big\}
\, ,
\\
&
\int_0^x d\tilde{x}
\tilde{x}
j_1(a\tilde{x})j_0(b\tilde{x}) y_0(c\tilde{x})
\nonumber\\
&
= 
-\frac{x}{a} j_0(ax) j_0(bx) y_0(cx)
- \frac{1}{2}x^2 j_1(ax) j_0(bx) y_0(cx)
+ \frac{b}{2a}x^2 j_0(ax) j_1(bx) y_0(cx)
+ \frac{c}{2a}x^2 j_0(ax) j_0(bx) y_1(cx)
\nonumber\\
&
\quad
+ \frac{1}{8a^2bc} \Big\{
\big( a^2-(b+c)^2 \big) {\rm Cin}[(a-b-c)x] - \big( a^2-(b-c)^2 \big) {\rm Cin}[(a+b-c)x]
\nonumber\\
&
\quad
\hspace{4em}
+ \big( a^2-(b-c)^2 \big) {\rm Cin}[(a-b+c)x] - \big( a^2-(b+c)^2 \big) {\rm Cin}[(a+b+c)x]
\Big\}
- \frac{1}{2ac}
\, ,
\end{align}
\begin{align}
&
\int_0^x d\tilde{x}
j_1(a\tilde{x})j_1(b\tilde{x}) j_0(c\tilde{x})
\nonumber\\
&
= 
- \frac{a^2+b^2}{3ab}x j_0(ax) j_0(bx) j_0(cx)
+ \frac{a^2-b^2}{8b}x^2 j_1(ax) j_0(bx) j_0(cx)
+ \frac{-a^2+b^2}{8a}x^2 j_0(ax) j_1(bx) j_0(cx)
\nonumber\\
&
\quad
+ \frac{c^2}{24b}x^2 j_1(ax) j_0(bx) j_0(cx)
+ \frac{c^2}{24a}x^2 j_0(ax) j_1(bx) j_0(cx)
+ \frac{c(5a^2+5b^2-c^2)}{24ab}x^2 j_0(ax) j_0(bx) j_1(cx)
\nonumber\\
&
\quad
- \frac{1}{3}x j_1(ax) j_1(bx) j_0(cx)
+ \frac{c}{12}x^2 j_1(ax) j_1(bx) j_1(cx)
\nonumber\\
&
\quad
+ \frac{1}{96a^2b^2c} \Big\{
(a-b-c)^2 (3a^2 + 6ab + 3b^2 - 2ac + 2bc - c^2) {\rm Si}[(a-b-c)x]
\nonumber\\
&
\quad
\hspace{5em}
- (a+b-c)^2 (3a^2 - 6ab + 3b^2 - 2ac - 2bc - c^2) {\rm Si}[(a+b-c)x]
\nonumber\\
&
\quad
\hspace{5em}
- (a-b+c)^2 (3a^2 + 6ab + 3b^2 + 2ac - 2bc - c^2) {\rm Si}[(a-b+c)x]
\nonumber\\
&
\quad
\hspace{5em}
+ (a+b+c)^2 (3a^2 - 6ab + 3b^2 + 2ac + 2bc - c^2) {\rm Si}[(a+b+c)x]
\Big\}
\, ,
\\
&
\int_0^x d\tilde{x}
j_1(a\tilde{x})j_1(b\tilde{x}) y_0(c\tilde{x})
\nonumber\\
&
= 
- \frac{a^2+b^2}{3ab}x j_0(ax) j_0(bx) y_0(cx)
+ \frac{a^2-b^2}{8b}x^2 j_1(ax) j_0(bx) y_0(cx)
+ \frac{-a^2+b^2}{8a}x^2 j_0(ax) j_1(bx) y_0(cx)
\nonumber\\
&
\quad
+ \frac{c^2}{24b}x^2 j_1(ax) j_0(bx) y_0(cx)
+ \frac{c^2}{24a}x^2 j_0(ax) j_1(bx) y_0(cx)
+ \frac{c(5a^2+5b^2-c^2)}{24ab}x^2 j_0(ax) j_0(bx) y_1(cx)
\nonumber\\
&
\quad
- \frac{1}{3}x j_1(ax) j_1(bx) y_0(cx)
+ \frac{c}{12}x^2 j_1(ax) j_1(bx) y_1(cx)
\nonumber\\
&
\quad
+ \frac{1}{96a^2b^2c} \Big\{
-(a-b-c)^2 (3a^2 + 6ab + 3b^2 - 2ac + 2bc - c^2) {\rm Cin}[(a-b-c)x]
\nonumber\\
&
\quad
\hspace{5em}
+ (a+b-c)^2 (3a^2 - 6ab + 3b^2 - 2ac - 2bc - c^2) {\rm Cin}[(a+b-c)x]
\nonumber\\
&
\quad
\hspace{5em}
- (a-b+c)^2 (3a^2 + 6ab + 3b^2 + 2ac - 2bc - c^2) {\rm Cin}[(a-b+c)x]
\nonumber\\
&
\quad
\hspace{5em}
+ (a+b+c)^2 (3a^2 - 6ab + 3b^2 + 2ac + 2bc - c^2) {\rm Cin}[(a+b+c)x]
\Big\}
- \frac{3a^2+3b^2+c^2}{24abc}
\, .
\end{align}


For the comoving gauge, we can find
\begin{align}
\int_0^x d\tilde{x} \tilde{x}^3 j_1(a\tilde{x}) j_0(b\tilde{x}) j_0(c\tilde{x})
& =
\frac{x}{4abc} \Big\{ j_0[(a-b-c)x] - j_0[(a+b-c)x] - j_0[(a-b+c)x] + j_0[(a+b+c)x] \Big\}
\nonumber\\
& \quad
+ \frac{1}{4a^2bc} \Big\{ {\rm Si}[(a-b-c)x] + {\rm Si}[(a+b-c)x] + {\rm Si}[(a-b+c)x] - {\rm Si}[(a+b+c)x] \Big\}
\, ,
\\
\int_0^x d\tilde{x} \tilde{x}^3 j_1(a\tilde{x}) j_0(b\tilde{x}) y_0(c\tilde{x})
& =
\frac{x}{4abc} \Big\{ -y_0[(a-b-c)x] + y_0[(a+b-c)x] - y_0[(a-b+c)x] + y_0[(a+b+c)x] \Big\}
\nonumber\\
& \quad
+ \frac{1}{4a^2bc} \Big\{ {\rm Cin}[(a-b-c)x] - {\rm Cin}[(a+b-c)x] + {\rm Cin}[(a-b+c)x] - {\rm Cin}[(a+b+c)x] \Big\}
\nonumber\\
&
\quad 
- \frac{a^2-b^2+c^2}{ac(a-b-c)(a+b-c)(a-b+c)(a+b+c)}
\, ,
\end{align}
\begin{align}
\int_0^x d\tilde{x} \tilde{x}^4 j_0(a\tilde{x}) j_0(b\tilde{x}) j_0(c\tilde{x})
& =
\frac{x^2}{4abc} \Big\{ -j_1[(a-b-c)x] + j_1[(a+b-c)x] + j_1[(a-b+c)x] - j_1[(a+b+c)x] \Big\}
\, ,
\\
\int_0^x d\tilde{x} \tilde{x}^4 j_0(a\tilde{x}) j_0(b\tilde{x}) y_0(c\tilde{x})
& =
\frac{x^2}{4abc} \Big\{ y_1[(a-b-c)x] - y_1[(a+b-c)x] + y_1[(a-b+c)x] - y_1[(a+b+c)x] \Big\}
\nonumber\\
&
\quad
+ \frac{2 \big( a^4 + b^4 - 3c^4 - 2a^2b^2 + 2a^2c^2 + 2b^2c^2 \big)}
{(a-b-c)^2(a+b-c)^2(a-b+c)^2(a+b+c)^2} 
\, ,
\end{align}
\begin{align}
\int_0^x d\tilde{x} \tilde{x}^4 j_1(a\tilde{x}) j_1(b\tilde{x}) j_0(c\tilde{x})
& =
\frac{x^2}{4abc} \Big\{ - y_0[(a-b-c)x] - y_0[(a+b-c)x] + y_0[(a-b+c)x] + y_0[(a+b+c)x] \Big\}
\nonumber\\
& \quad
+ \frac{x^2}{4a^2b^2c} \Big\{ 
\big( a^2 - 3ab + b^2 - ac + bc \big) \Big[ j_1[(a-b-c)x] - y_0[(a-b-c)x] \Big]
\nonumber\\
&
\hspace{5.5em}
- \big( a^2 + 3ab + b^2 - ac - bc \big) \Big[ j_1[(a+b-c)x] - y_0[(a+b-c)x] \Big]
\nonumber\\
&
\hspace{5.5em}
- \big( a^2 - 3ab + b^2 + ac - bc \big) \Big[ j_1[(a-b+c)x] - y_0[(a-b+c)x] \Big]
\nonumber\\
&
\hspace{5.5em}
+ \big( a^2+ 3ab + b^2 + ac + bc \big) \Big[ j_1[(a+b+c)x] - y_0[(a+b+c)x] \Big]
\Big\}
\nonumber\\
&
\quad
+ \frac{1}{4a^2b^2c} \Big\{ - {\rm Si}[(a-b-c)x] + {\rm Si}[(a+b-c)x] + {\rm Si}[(a-b+c)x] - {\rm Si}[(a+b+c)x] \Big\}
\, ,
\\
\int_0^x d\tilde{x} \tilde{x}^4 j_1(a\tilde{x}) j_1(b\tilde{x}) y_0(c\tilde{x})
& =
- \frac{x^2}{4abc} \Big\{ j_0[(a-b-c)x] + j_0[(a+b-c)x] + j_0[(a-b+c)x] + j_0[(a+b+c)x] \Big\}
\nonumber\\
& \quad
+ \frac{x^2}{4a^2b^2c} \Big\{
- \big( a^2 - 3ab + b^2 - ac + bc \big) \Big[ y_1[(a-b-c)x] + j_0[(a-b-c)x] \Big]
\nonumber\\
&
\hspace{5.5em}
+ \big( a^2 + 3ab + b^2 - ac - bc \big) \Big[ y_1[(a+b-c)x] + j_0[(a+b-c)x] \Big]
\nonumber\\
&
\hspace{5.5em}
- \big( a^2 - 3ab + b^2 + ac - bc \big) \Big[ y_1[(a-b+c)x] + j_0[(a-b+c)x] \Big]
\nonumber\\
&
\hspace{5.5em}
+ \big( a^2+ 3ab + b^2 + ac + bc \big) \Big[ y_1[(a+b+c)x] + j_0[(a+b+c)x] \Big]
\Big\}
\nonumber\\
& 
\quad
+ \frac{a^6 + (b^2-c^2)^3 - a^4(b^2+3c^2) + a^2(-b^4+14b^2c^2+3c^4)}
{abc(a-b-c)^2(a+b-c)^2(a-b+c)^2(a+b+c)^2}
\, .
\end{align}

\subsection{Integrals for scalar-tensor induced GW during RD}
\label{app:RDst}

We find
\begin{align}
&
\int_0^x d\tilde{x} \tilde{x} j_1(a\tilde{x}) j_1(b\tilde{x}) j_0(c\tilde{x})
\nonumber\\
&
= 
\frac{1}{6ab} j_0(ax) j_0(bx) j_0 (cx)
- \frac{a}{3b}x^2 j_0(ax) j_0(bx) j_0 (cx)
- \frac{b}{3a}x^2 j_0(ax) j_0(bx) j_0 (cx)
+ \frac{c^2}{6ab}x^2 j_0(ax) j_0(bx) j_0 (cx)
\nonumber\\
&
\quad
- \frac{1}{6b}x j_1(ax) j_0(bx) j_0 (cx)
- \frac{1}{6a}x j_0(ax) j_1(bx) j_0 (cx)
- \frac{1}{3}x^2 j_1(ax) j_1(bx) j_0 (cx)
- \frac{c}{6ab}x j_0(ax) j_0(bx) j_1(cx)
\nonumber\\
&
\quad
+ \frac{c}{6b}x^2 j_1(ax) j_0(bx) j_1(cx)
+ \frac{c}{6a}x^2 j_0(ax) j_1(bx) j_1(cx)
\nonumber\\
&
\quad
+ \frac{1}{24a^2b^2c} \Big\{
(a-b-c) \big( 2a^2 + 2ab + 2b^2 - ac + bc - c^2 \big) {\rm Cin}[(a-b-c)x]
\nonumber\\
&
\quad
\hspace{5em}
- (a+b-c) \big( 2a^2 - 2ab + 2b^2 - ac - bc - c^2 \big) {\rm Cin}[(a+b-c)x]
\nonumber\\
&
\quad
\hspace{5em}
- (a-b+c) \big( 2a^2 + 2ab + 2b^2 + ac - bc - c^2 \big) {\rm Cin}[(a-b+c)x]
\nonumber\\
&
\quad
\hspace{5em}
+ (a+b+c) \big( 2a^2 - 2ab + 2b^2 + ac + bc - c^2 \big) {\rm Cin}[(a+b+c)x]
\Big\}
- \frac{1}{6ab}
\, ,
\end{align}
\begin{align}
&
\int_0^x d\tilde{x} \tilde{x} j_1(a\tilde{x}) j_1(b\tilde{x}) y_0(c\tilde{x})
\nonumber\\
&
= 
\frac{1}{6ab} j_0(ax) j_0(bx) y_0(cx)
- \frac{a}{3b}x^2 j_0(ax) j_0(bx) y_0(cx)
- \frac{b}{3a}x^2 j_0(ax) j_0(bx) y_0(cx)
+ \frac{c^2}{6ab}x^2 j_0(ax) j_0(bx) y_0(cx)
\nonumber\\
&
\quad
- \frac{1}{6b}x j_1(ax) j_0(bx) y_0(cx)
- \frac{1}{6a}x j_0(ax) j_1(bx) y_0(cx)
- \frac{1}{3}x^2 j_1(ax) j_0(bx) y_0(cx)
- \frac{c}{6ab}x j_0(ax) j_0(bx) y_1(cx)
\nonumber\\
&
\quad
+ \frac{c}{6b}x^2 j_1(ax) j_0(bx) y_1(cx)
+ \frac{c}{6a}x^2 j_0(ax) j_1(bx) y_1(cx)
\nonumber\\
&
\quad
+ \frac{1}{24a^2b^2c} \Big\{
(a-b-c) \big( 2a^2 + 2ab + 2b^2 - ac + bc - c^2 \big) {\rm Si}[(a-b-c)x]
\nonumber\\
&
\quad
\hspace{5em}
- (a+b-c) \big( 2a^2 - 2ab + 2b^2 - ac - bc - c^2 \big) {\rm Si}[(a+b-c)x]
\nonumber\\
&
\quad
\hspace{5em}
+ (a-b+c) \big( 2a^2 + 2ab + 2b^2 + ac - bc - c^2 \big) {\rm Si}[(a-b+c)x]
\nonumber\\
&
\quad
\hspace{5em}
- (a+b+c) \big( 2a^2 - 2ab + 2b^2 + ac + bc - c^2 \big) {\rm Si}[(a+b+c)x]
\Big\}
\, ,
\end{align}
\begin{align}
&
\int_0^x d\tilde{x} \tilde{x} j_1(a\tilde{x}) j_0(b\tilde{x}) j_0(c\tilde{x})
\nonumber\\
&
=
- \frac{1}{a}x j_0(ax) j_0(bx) j_0(cx)
- \frac{1}{2}x^2 j_1(ax) j_0(bx) j_0(cx)
+ \frac{b}{2a}x^2 j_0(ax) j_1(bx) j_0(cx)
+ \frac{c}{2a}x^2 j_0(ax) j_0(bx) j_1(cx)
\nonumber\\
&
\quad
+ \frac{1}{8a^2bc} \Big\{
- \big[ a^2 - (b+c)^2 \big] {\rm Si}[(a-b-c)x]
+ \big[ a^2 - (b-c)^2 \big] {\rm Si}[(a+b-c)x]
\nonumber\\
&
\quad
\hspace{4em}
+ \big[ a^2 - (b-c)^2 \big] {\rm Si}[(a-b+c)x]
- \big[ a^2 - (b+c)^2 \big] {\rm Si}[(a+b+c)x]
\Big\}
\, ,
\end{align}
\begin{align}
&
\int_0^x d\tilde{x} \tilde{x} j_1(a\tilde{x}) j_0(b\tilde{x}) y_0(c\tilde{x})
\nonumber\\
&
=
- \frac{1}{a}x j_0(ax) j_0(bx) y_0(cx)
- \frac{1}{2}x^2 j_1(ax) j_0(bx) y_0(cx)
+ \frac{b}{2a}x^2 j_0(ax) j_1(bx) y_0(cx)
+ \frac{c}{2a}x^2 j_0(ax) j_0(bx) y_1(cx)
\nonumber\\
&
\quad
+ \frac{1}{8a^2bc} \Big\{
\big[ a^2 - (b+c)^2 \big] {\rm Cin}[(a-b-c)x]
- \big[ a^2 - (b-c)^2 \big] {\rm Cin}[(a+b-c)x]
\nonumber\\
&
\quad
\hspace{4em}
+ \big[ a^2 - (b-c)^2 \big] {\rm Cin}[(a-b+c)x]
- \big[ a^2 - (b+c)^2 \big] {\rm Cin}[(a+b+c)x]
\Big\}
- \frac{1}{2ac} 
\, ,
\end{align}
\begin{align}
&
\int_0^x d\tilde{x} j_2(a\tilde{x}) j_0(b\tilde{x}) j_0(c\tilde{x})
\nonumber\\
&
= 
\frac{b^2}{2a^2}x j_0(ax) j_0(bx) j_0(cx)
+ \frac{c^2}{2a^2}x j_0(ax) j_0(bx) j_0(cx)
- \frac{3}{2a} j_1(ax) j_0(bx) j_0(cx)
- \frac{a}{8}x^2 j_1(ax) j_0(bx) j_0(cx) 
\nonumber\\
&
\quad
+ \frac{b^2}{8a}x^2 j_1(ax) j_0(bx) j_0(cx)
+ \frac{c^2}{8a}x^2 j_1(ax) j_0(bx) j_0(cx)
+ \frac{b}{8}x^2 j_0(ax) j_1(bx) j_0(cx)
- \frac{b^3}{8a^2}x^2 j_0(ax) j_1(bx) j_0(cx)
\nonumber\\
&
\quad
- \frac{3bc^2}{8a^2}x^2 j_0(ax) j_1(bx) j_0(cx)
+ \frac{b}{2a}x j_1(ax) j_1(bx) j_0(cx)
+ \frac{c}{8}x^2 j_0(ax) j_0(bx) j_1(cx)
- \frac{3b^2c}{8a^2}x^2 j_0(ax) j_0(bx) j_1(cx)
\nonumber\\
&
\quad
- \frac{c^3}{8a^2}x^2 j_0(ax) j_0(bx) j_1(cx)
+ \frac{c}{2a}x j_1(ax) j_0(bx) j_1(cx)
- \frac{bc}{4a}x^2 j_1(ax) j_1(bx) j_1(cx)
\nonumber\\
&
\quad
+ \frac{1}{32a^3bc} \Big\{
- \big[ a^2-(b+c)^2 \big]^2 {\rm Si}[(a-b-c)x]
+ \big[ a^2-(b-c)^2 \big]^2 {\rm Si}[(a+b-c)x]
\nonumber\\
&
\quad
\hspace{5em}
+ \big[ a^2-(b-c)^2 \big]^2 {\rm Si}[(a-b+c)x]
- \big[ a^2-(b+c)^2 \big]^2 {\rm Si}[(a+b+c)x]
\Big\}
\, ,
\end{align}
\begin{align}
&
\int_0^x d\tilde{x} j_2(a\tilde{x}) j_0(b\tilde{x}) y_0(c\tilde{x})
\nonumber\\
&
= 
\frac{b^2}{2a^2}x j_0(ax) j_0(bx) y_0(cx)
+ \frac{c^2}{2a^2}x j_0(ax) j_0(bx) y_0(cx)
- \frac{3}{2a} j_1(ax) j_0(bx) y_0(cx)
- \frac{a}{8}x^2 j_1(ax) j_0(bx) y_0(cx)
\nonumber\\
&
\quad
+ \frac{b^2}{8a}x^2 j_1(ax) j_0(bx) y_0(cx)
+ \frac{c^2}{8a}x^2 j_1(ax) j_0(bx) y_0(cx)
+ \frac{b}{8}x^2 j_0(ax) j_1(bx) y_0(cx)
- \frac{b^3}{8a^2}x^2 j_0(ax) j_1(bx) y_0(cx)
\nonumber\\
&
\quad
- \frac{3bc^2}{8a^2}x^2 j_0(ax) j_1(bx) y_0(cx)
+ \frac{b}{2a}x j_1(ax) j_1(bx) y_0(cx)
+ \frac{c}{8}x^2 j_0(ax) j_0(bx) y_1(cx)
- \frac{3b^2c}{8a^2}x^2 j_0(ax) j_0(bx) y_1(cx)
\nonumber\\
&
\quad
- \frac{c^3}{8a^2}x^2 j_0(ax) j_0(bx) y_1(cx)
+ \frac{c}{2a}x j_1(ax) j_0(bx) y_1(cx)
- \frac{bc}{4a^2}x^2 j_1(ax) j_1(bx) y_1(cx)
\nonumber\\
&
\quad
+ \frac{1}{32a^3bc} \Big\{
\big[ a^2-(b+c)^2 \big]^2 {\rm Cin}[(a-b-c)x]
- \big[ a^2-(b-c)^2 \big]^2 {\rm Cin}[(a+b-c)x]
\nonumber\\
&
\quad
\hspace{5em}
+ \big[ a^2-(b-c)^2 \big]^2 {\rm Cin}[(a-b+c)x]
- \big[ a^2-(b+c)^2 \big]^2 {\rm Cin}[(a+b+c)x]
\Big\}
- \frac{5}{24c} + \frac{b^2}{8a^2c} + \frac{3c}{8a^2}
\, .
\end{align}

\subsection{Integrals for tensor-tensor induced GWs during RD}

We can perform the integrals analytically to find
\begin{align}
&
\int_0^x d\tilde{x} \tilde{x}^2 j_1(a\tilde{x})j_1(b\tilde{x})j_0(c\tilde{x})
\nonumber\\
& =
\frac{c}{2ab}x^2 j_0(ax) j_0(bx) j_1(cx) 
- \frac{1}{2a} x^2 j_0(ax) j_1(bx) j_0(cx)
- \frac{1}{2b} x^2 j_1(ax) j_0(bx) j_0(cx)
\nonumber\\
&
\quad
- \frac{a^2+b^2-c^2}{8a^2b^2c} \Big\{
{\rm Si}\big[ (a-b-c)x\big] - {\rm Si}\big[ (a+b-c)x\big] - {\rm Si}\big[ (a-b+c)x\big] + {\rm Si}\big[ (a+b+c)x\big]
\Big\}
\, ,
\\
&
\int_0^x d\tilde{x} \tilde{x}^2 j_1(a\tilde{x})j_1(b\tilde{x})y_0(c\tilde{x})
\nonumber\\
& =
\frac{c}{2ab}x^2 j_0(ax) j_0(bx) y_1(cx)
- \frac{1}{2a}x^2 j_0(ax) j_1(bx) y_0(cx)
- \frac{1}{2b}x^2 j_1(ax) j_0(bx) y_0(cx)
+ \frac{1}{2abc}
\nonumber\\
&
\quad
+ \frac{a^2+b^2-c^2}{8a^2b^2c} \Big\{
{\rm Cin}\big[(a-b-c)x\big] - {\rm Cin}\big[(a+b-c)x\big] + {\rm Cin}\big[(a-b+c)x\big] - {\rm Cin}\big[(a+b+c)x\big]
\Big\}
\, .
\end{align}
Thus,
\begin{align}
\label{eq:RDttF}
&
\int_0^x d\tilde{x} \tilde{x}^2
j_1(a\tilde{x}) j_1(b\tilde{x})
\Big[ j_0(\tilde{x})y_0(x) - j_0(x)y_0(\tilde{x}) \Big]
\equiv 
\frac{1}{2ab} {F}_\text{RD}(a,b,x)
\nonumber\\
& =
\frac{1}{2ab} \Bigg[ j_0(ax) j_0(bx) -  j_0(x)
+ \frac{a^2+b^2-1}{4ab} \bigg\{
\Big( -{\rm Si}\big[ (a-b-1)x\big] y_0(x) - {\rm Cin}\big[(a-b-1)x\big] j_0(x) \Big) 
\nonumber\\
&
\hspace{19em}
- \Big( -{\rm Si}\big[ (a+b-1)x\big] y_0(x) - {\rm Cin}\big[(a+b-1)x\big] j_0(x) \Big) 
\nonumber\\
&
\hspace{19em}
- \Big( -{\rm Si}\big[ (a-b+1)x\big] y_0(x) + {\rm Cin}\big[(a-b+1)x\big] j_0(x) \Big) 
\nonumber\\
&
\hspace{19em}
+ \Big( -{\rm Si}\big[ (a+b+1)x\big] y_0(x) + {\rm Cin}\big[(a+b+1)x\big] j_0(x) \Big) 
\bigg\}
\Bigg]
\, ,
\\
\label{eq:RDttG}
&
\int_0^x d\tilde{x} \tilde{x}^2
j_0(a\tilde{x}) j_0(b\tilde{x})
\Big[ j_0(\tilde{x})y_0(x) - j_0(x)y_0(\tilde{x}) \Big] 
\equiv
\frac{1}{2} {G}_\text{RD}(a,b,x)
\nonumber\\
& =
\frac{1}{4ab} \bigg\{
\Big( -{\rm Si}\big[ (a-b-1)x\big] y_0(x) - {\rm Cin}\big[(a-b-1)x\big] j_0(x) \Big) 
+ \Big( {\rm Si}\big[ (a+b-1)x\big] y_0(x) + {\rm Cin}\big[(a+b-1)x\big] j_0(x) \Big) 
\nonumber\\
&
\hspace{3em}
+ \Big( {\rm Si}\big[ (a-b+1)x\big] y_0(x) - {\rm Cin}\big[(a-b+1)x\big] j_0(x) \Big) 
+ \Big( - {\rm Si}\big[ (a+b+1)x\big] y_0(x) + {\rm Cin}\big[(a+b+1)x\big] j_0(x) \Big) 
\bigg\}
\, .
\end{align}
Note that
\begin{align}
\frac{a^2+b^2-1}{2} {G}_\text{RD}(a,b,x)
=
{F}_\text{RD}(a,b,x) - j_0(ax)j_0(bx) + j_0(x)
\, .
\end{align}


\begin{thebibliography}{99}





\bibitem[Abbott et al.(2016a)]{2016PhRvL.116f1102A} Abbott, B.~P., Abbott, R., Abbott, T.~D., et al.\ 2016, \prl, 116, 061102





\bibitem[Abbott et al.(2016b)]{2016PhRvL.116x1103A} Abbott, B.~P., Abbott, R., Abbott, T.~D., et al.\ 2016, \prl, 116, 241103





\bibitem[Abbott et al.(2017a)]{2017PhRvL.118v1101A} Abbott, B.~P., Abbott, R., Abbott, T.~D., et al.\ 2017, \prl, 118, 221101





\bibitem[Abbott et al.(2017b)]{2017PhRvL.119n1101A} Abbott, B.~P., Abbott, R., Abbott, T.~D., et al.\ 2017, \prl, 119, 141101





\bibitem[Abbott et al.(2017c)]{2017PhRvL.119p1101A} Abbott, B.~P., Abbott, R., Abbott, T.~D., et al.\ 2017, \prl, 119, 161101

  
  
  
  
\bibitem[Abbott et al.(2017d)]{2017ApJ...851L..35A} Abbott, B.~P., Abbott, R., Abbott, T.~D., et al.\ 2017, \apjl, 851, L35





\bibitem[Alabidi et al.(2012)]{2012JCAP...09..017A} Alabidi, L., Kohri, K., Sasaki, M., et al.\ 2012, \jcap, 2012, 017





\bibitem[Alabidi et al.(2013)]{2013JCAP...05..033A} Alabidi, L., Kohri, K., Sasaki, M., et al.\ 2013, \jcap, 2013, 033





\bibitem[Ananda et al.(2007)]{2007PhRvD..75l3518A} Ananda, K.~N., Clarkson, C., \& Wands, D.\ 2007, \prd, 75, 123518





\bibitem[Arnowitt et al.(2008)]{2008GReGr..40.1997A} Arnowitt, R., Deser, S., \& Misner, C.~W.\ 2008, General Relativity and Gravitation, 40, 1997





\bibitem[Arroja et al.(2009)]{2009PhRvD..80l3526A} Arroja, F., Assadullahi, H., Koyama, K., et al.\ 2009, \prd, 80, 123526





\bibitem[Assadullahi, \& Wands(2009)]{2009PhRvD..79h3511A} Assadullahi, H., \& Wands, D.\ 2009, \prd, 79, 083511





\bibitem[Assadullahi, \& Wands(2010)]{2010PhRvD..81b3527A} Assadullahi, H., \& Wands, D.\ 2010, \prd, 81, 023527





\bibitem[Baumann et al.(2007)]{2007PhRvD..76h4019B} Baumann, D., Steinhardt, P., Takahashi, K., et al.\ 2007, \prd, 76, 084019





\bibitem[BICEP2 Collaboration et al.(2016)]{2016PhRvL.116c1302B} BICEP2 Collaboration, Keck Array Collaboration, Ade, P. A. R., et al.\ 2016, \prl, 116, 031302





\bibitem[Brandenberger et al.(1986)]{1986NuPhB.277..605B} Brandenberger, R.~H., Albrecht, A., \& Turok, N.\ 1986, Nuclear Physics B, 277, 605




\bibitem[Brandenberger et al.(2007)]{2007PhRvL..98w1302B} Brandenberger, R.~H., Nayeri, A., Patil, S.~P., et al.\ 2007, \prl, 98, 231302




\bibitem[Cai et al.(2015)]{2015NuPhB.900..517C} Cai, Y.-F., Gong, J.-O., Pi, S., et al.\ 2015, Nuclear Physics B, 900, 517




\bibitem[De Luca et al.(2020)]{2020JCAP...03..014D} De Luca, V., Franciolini, G., Kehagias, A., et al.\ 2020, \jcap, 2020, 014





\bibitem[Espinosa et al.(2018)]{2018JCAP...09..012E} Espinosa, J.~R., Racco, D., \& Riotto, A.\ 2018, \jcap, 2018, 012





\bibitem[Gong, \& Stewart(2002)]{2002PhLB..538..213G} Gong, J.-O., \& Stewart, E.~D.\ 2002, Physics Letters B, 538, 213




\bibitem[Gong(2014)]{2014JCAP...07..022G} Gong, J.-O.\ 2014, \jcap, 2014, 022





\bibitem[Gong et al.(2017)]{2017JCAP...10..027G} Gong, J.-O., Hwang, J.-. chan ., Noh, H., et al.\ 2017, \jcap, 2017, 027





\bibitem[Hogan(1986)]{1986MNRAS.218..629H} Hogan, C.~J.\ 1986, \mnras, 218, 629





\bibitem[Hwang(1994)]{1994ApJ...427..533H} Hwang, J.-C.\ 1994, \apj, 427, 533





\bibitem[Hwang, \& Noh(2007)]{2007PhRvD..76j3527H} Hwang, J.-C., \& Noh, H.\ 2007, \prd, 76, 103527





\bibitem[Hwang et al.(2012)]{2012ApJ...752...50H} Hwang, J.-. chan ., Noh, H., \& Gong, J.-O.\ 2012, \apj, 752, 50





\bibitem[Hwang et al.(2017)]{2017ApJ...842...46H} Hwang, J.-. chan ., Jeong, D., \& Noh, H.\ 2017, \apj, 842, 46






\bibitem[Inomata et al.(2019a)]{2019JCAP...10..071I} Inomata, K., Kohri, K., Nakama, T., et al.\ 2019, \jcap, 2019, 071





\bibitem[Inomata et al.(2019b)]{2019PhRvD.100d3532I} Inomata, K., Kohri, K., Nakama, T., et al.\ 2019, \prd, 100, 043532




\bibitem[Inomata \& Terada(2020)]{2020PhRvD.101b3523I} Inomata, K. \& Terada, T.\ 2020, \prd, 101, 023523




\bibitem[Jain et al.(2009)]{2009JCAP...01..009J} Jain, R.~K., Chingangbam, P., Gong, J.-O., et al.\ 2009, \jcap, 2009, 009




\bibitem[Jain et al.(2010)]{2010PhRvD..82b3509J} Jain, R.~K., Chingangbam, P., Sriramkumar, L., et al.\ 2010, \prd, 82, 023509





\bibitem[Jedamzik et al.(2010)]{2010JCAP...04..021J} Jedamzik, K., Lemoine, M., \& Martin, J.\ 2010, \jcap, 2010, 021





\bibitem[Khlebnikov, \& Tkachev(1997)]{1997PhRvD..56..653K} Khlebnikov, S., \& Tkachev, I.\ 1997, \prd, 56, 653




\bibitem[Kobayashi et al.(2010)]{2010PhRvL.105w1302K} Kobayashi, T., Yamaguchi, M., \& Yokoyama, J.\ 2010, \prl, 105, 231302





\bibitem[Kohri, \& Terada(2018)]{2018PhRvD..97l3532K} Kohri, K., \& Terada, T.\ 2018, \prd, 97, 123532





\bibitem[Mollerach et al.(2004)]{2004PhRvD..69f3002M} Mollerach, S., Harari, D., \& Matarrese, S.\ 2004, \prd, 69, 063002





\bibitem[Mukhanov(2005)]{2005pfc..book.....M} Mukhanov, V.\ 2005, Physical Foundations of Cosmology, by Viatcheslav Mukhanov, pp. 442. Cambridge University Press, November 2005. 




\bibitem[Mukhanov, \& Vikman(2006)]{2006JCAP...02..004M} Mukhanov, V., \& Vikman, A.\ 2006, \jcap, 2006, 004




\bibitem[Mylova et al.(2018)]{2018JCAP...12..024M} Mylova, M., {\"O}zsoy, O., Parameswaran, S., et al.\ 2018, \jcap, 2018, 024




\bibitem[Nakama \& Suyama(2015)]{2015PhRvD..92l1304N} Nakama, T. \& Suyama, T.\ 2015, \prd, 92, 121304




\bibitem[Nakama \& Suyama(2016)]{2016PhRvD..94d3507N} Nakama, T. \& Suyama, T.\ 2016, \prd, 94, 043507





\bibitem[Noh, \& Hwang(2004)]{2004PhRvD..69j4011N} Noh, H., \& Hwang, J.-C.\ 2004, \prd, 69, 104011





\bibitem[Pi et al.(2019)]{2019JCAP...06..049P} Pi, S., Sasaki, M., \& Zhang, Y.-. li .\ 2019, \jcap, 2019, 049





\bibitem[Stewart, \& Gong(2001)]{2001PhLB..510....1S} Stewart, E.~D., \& Gong, J.-O.\ 2001, Physics Letters B, 510, 1




\bibitem[Tomikawa \& Kobayashi(2020)]{2020PhRvD.101h3529T} Tomikawa, K. \& Kobayashi, T.\ 2020, \prd, 101, 083529





\bibitem[Vachaspati, \& Vilenkin(1985)]{1985PhRvD..31.3052V} Vachaspati, T., \& Vilenkin, A.\ 1985, \prd, 31, 3052





\bibitem[Witten(1984)]{1984PhRvD..30..272W} Witten, E.\ 1984, \prd, 30, 272




  
\bibitem[Yoo, \& Gong(2016)]{2016JCAP...07..017Y} Yoo, J., \& Gong, J.-O.\ 2016, \jcap, 2016, 017




\bibitem[Yuan et al.(2020)]{2020PhRvD.101f3018Y} Yuan, C., Chen, Z.-C., \& Huang, Q.-G.\ 2020, \prd, 101, 063018






\end{thebibliography}
\end{document}